\documentclass[runningheads]{llncs}
\usepackage{graphicx}

\usepackage{tikz}
\usepackage{comment}
\usepackage{amsmath,amssymb} 
\usepackage{color}
\usepackage{hyperref}
\usepackage{booktabs}
\usepackage{multirow}
\usepackage{graphics}
\usepackage{graphicx, caption, subcaption}

\hypersetup{
    colorlinks=true,
    linkcolor=magenta,
    filecolor=blue,      
    urlcolor=cyan,
    }

\newcommand{\etal}{\textit{et al.}}
\newcommand{\eg}{\textit{e.g.}}
\newcommand{\ie}{\textit{i.e.}}
\newcommand{\bm}[1]{\boldsymbol{#1}}

\usepackage[accsupp]{axessibility}  

\begin{document}
\pagestyle{headings}
\mainmatter
\def\ECCVSubNumber{10}  

\title{MIPI 2022 Challenge on Under-Display Camera Image Restoration: Methods and Results} 

\titlerunning{MIPI 2022 Challenge on UDC Image Restoration}
%
\author{Ruicheng Feng \and
Chongyi Li \and
Shangchen Zhou \and
Wenxiu Sun \and
Qingpeng Zhu \and
Jun Jiang \and
Qingyu Yang \and
Chen Change Loy \and
Jinwei Gu \and
Yurui Zhu \and
Xi Wang \and
Xueyang Fu \and
Xiaowei Hu \and
Jinfan Hu \and
Xina Liu \and
Xiangyu Chen \and
Chao Dong \and
Dafeng Zhang \and
Feiyu Huang \and
Shizhuo Liu \and
Xiaobing Wang \and
Zhezhu Jin \and
Xuhao Jiang \and
Guangqi Shao \and
Xiaotao Wang \and
Lei Lei \and
Zhao Zhang \and
Suiyi Zhao \and
Huan Zheng \and
Yangcheng Gao \and
Yanyan Wei \and
Jiahuan Ren \and
Tao Huang \and
Zhenxuan Fang \and
Mengluan Huang \and
Junwei Xu \and
Yong Zhang \and
Yuechi Yang \and
Qidi Shu \and
Zhiwen Yang \and
Shaocong Li \and
Mingde Yao \and
Ruikang Xu \and
Yuanshen Guan \and
Jie Huang \and
Zhiwei Xiong \and
Hangyan Zhu \and
Ming Liu \and
Shaohui Liu \and
Wangmeng Zuo \and
Zhuang Jia \and
Binbin Song \and
Ziqi Song \and
Guiting Mao \and
Ben Hou \and
Zhimou Liu \and
Yi Ke \and
Dengpei Ouyang \and
Dekui Han \and
Jinghao Zhang \and
Qi Zhu \and
Naishan Zheng \and
Feng Zhao \and
Wu Jin \and
Marcos Conde \and
Sabari Nathan \and
Radu Timofte \and
Tianyi Xu \and
Jun Xu \and
Hrishikesh P.S. \and
Densen Puthussery \and
Jiji C.V. \and
Biao Jiang \and
Yuhan Ding \and
Wanzhang Li \and
Xiaoyue Feng \and
Sijing Chen \and
Tianheng Zhong \and
Jiyang Lu \and
Hongming Chen \and
Zhentao Fan \and
Xiang Chen
\institute{~}
}
\authorrunning{R. Feng et al.}

\maketitle
\begin{sloppypar}
\let\thefootnote\relax\footnotetext{\tiny Ruicheng Feng$^{1}$ (\email{ruicheng002@ntu.edu.sg}), Chongyi Li$^{1}$ (\email{chongyi.li@ntu.edu.sg}), Shangchen Zhou$^{1}$, Wenxiu Sun$^{3,4}$, Qingpeng Zhu$^{4}$, Jun Jiang$^{2}$, 
Qingyu Yang$^{2}$, Chen Change Loy$^{1}$, Jinwei Gu$^{2,3}$ are the MIPI 2022 challenge organizers
($^{1}$Nanyang Technological University, $^{2}$SenseBrain, $^{3}$Shanghai AI Laboratory, $^{4}$SenseTime Research and Tetras.AI). The other authors participated in the challenge. Please refer to Appendix for details.\\ 
MIPI 2022 challenge website: \url{http://mipi-challenge.org}}

\begin{abstract}
Developing and integrating advanced image sensors with novel algorithms in camera systems are prevalent with the increasing demand for computational photography and imaging on mobile platforms.
However, the lack of high-quality data for research and the rare opportunity for in-depth exchange of views from industry and academia constrain the development of mobile intelligent photography and imaging (MIPI).
To bridge the gap, we introduce the first MIPI challenge including five tracks focusing on novel image sensors and imaging algorithms.
In this paper, we summarize and review the Under-Display Camera (UDC) Image Restoration track on MIPI 2022.
In total, 167 participants were successfully registered, and 19 teams submitted results in the final testing phase.
The developed solutions in this challenge achieved state-of-the-art performance on Under-Display Camera Image Restoration.
A detailed description of all models developed in this challenge is provided in this paper.
More details of this challenge and the link to the dataset can be found at \href{https://github.com/mipi-challenge/MIPI2022}{https://github.com/mipi-challenge/MIPI2022}.

\end{abstract}

\section{Introduction}
The demand for smartphones with full-screen displays has drawn interest from manufacturers in a newly-defined imaging system, Under-Display Camera (UDC). In addition, it also demonstrates practical applicability in other scenarios, \eg, for videoconferencing with a more natural gaze focus as cameras are placed at the center of the displays.

UDC is an imaging system whose camera is placed underneath a display.
However, widespread commercial productions of UDC are prevented by poor imaging quality caused by diffraction artifacts.
Such artifacts are unique to UDC, caused by the gaps between display pixels that act as an aperture and induce diffraction artifacts in the captured image.
Typical diffraction artifacts include flare, saturated blobs, blur, haze, and noise. 
Therefore, while bringing a better user experience, UDC may sacrifice image quality, and affect other downstream vision tasks.
The complex and diverse distortions make the reconstruction problem extremely challenging.
Zhou \etal \cite{zhou2021image,zhou2020udc} pioneered the attempt of the UDC image restoration and proposed a Monitor Camera Imaging System (MCIS) to capture paired data.
However, their work only simulated incomplete degradation. To alleviate this problem, Feng \etal \cite{feng2021removing} reformulated the image formation model and synthesized the UDC image by considering the diffraction flare of the saturated region in the high-dynamic-range (HDR) images.
This challenge is based on the dataset proposed in \cite{feng2021removing}, and aims to restore UDC images with complicated degradations. More details will be discussed in the following sections.

We hold this image restoration challenge in conjunction with MIPI Challenge which will be held on ECCV 2022. We are seeking an efficient and high-performance image restoration algorithm to be used for recovering under-display camera images. MIPI 2022 consists of five competition tracks:

\begin{itemize}
    \item RGB+ToF Depth Completion uses sparse and noisy ToF depth measurements with RGB images to obtain a complete depth map.
    \item Quad-Bayer Re-mosaic converts Quad-Bayer RAW data into Bayer format so that it can be processed by standard ISPs.
    \item RGBW Sensor Re-mosaic converts RGBW RAW data into Bayer format so that it can be processed by standard ISPs.
    \item RGBW Sensor Fusion fuses Bayer data and monochrome channel data into Bayer format to increase SNR and spatial resolution.
    \item Under-Display Camera Image Restoration improves the visual quality of images captured by a new imaging system equipped with an under-display camera.
\end{itemize}

\section{MIPI 2022 Under-Display Camera Image Restoration}
To facilitate the development of efficient and high-performance UDC image restoration solutions, we provide a high-quality dataset to be used for training and testing and a set of evaluation metrics that can measure the performance of developed solutions.
This challenge aims to advance research on UDC image restoration.

\subsection{Datasets}
The dataset is collected and synthesized using a model-based simulation pipeline as introduced in \cite{feng2021removing}.
The training split contains 2016 pairs of $800\times800\times3$ images. Image values are ranging from [0, 500] and constructed in `.npy' form.
The validation set is a subset of the testing set in \cite{feng2021removing}, and contains 40 pairs of images.
The testing set consists of another 40 pairs of images. The input images from the validation set and testing set are provided and the Ground Truth data are not available to participants.
Both input and Ground Truth data are high dynamic range. For evaluation, all measurements are computed in tone-mapped images (Modified Reinhard). The tone mapping operation can be expressed as $f(x) = \frac{x}{x+0.25}$.

\subsection{Evaluation}
The evaluation measures the objective fidelity and the perceptual quality of the UDC images with reference ground truth images.
We use the standard Peak Signal To Noise Ratio (PSNR) and the Structural Similarity (SSIM) index as often employed in the literature. In addition, Learned Perceptual Image Patch Similarity (LPIPS) \cite{zhang2018unreasonable} will be used as a complement. All measurements are computed in tone-mapped images (Modified Reinhard \cite{reinhard2002photographic}).
For the final ranking, we choose PSNR as the main measure, yet the top-ranked solutions are expected to also achieve above-average performance on SSIM and LPIPS.
For the dataset we report the average results over all the processed images.

\subsection{Challenge Phase}
The challenge consisted of the following phases:
\begin{enumerate}
    \item Development: The registered participants get access to the data and baseline code, and are able to train the models and evaluate their running time locally.
    \item Validation: The participants can upload their models to the remote server to check the fidelity scores on the validation dataset, and to compare their results on the validation leaderboard.
    \item Testing: The participants submit their final results, code, models, and factsheets.
\end{enumerate}

\section{Challenge Results}
Among $167$ registered participants, $19$ teams successfully submitted their results, code, and factsheets in the final test phase.
Table \ref{tab:result} reports the final test results and rankings of the teams. 
The methods evaluated in Table \ref{tab:result} are briefly
described in Section \ref{sec:methods} and the team members are listed in Appendix.
We have the following observations.
First, the USTC\_WXYZ team is the first place winner of this challenge, while XPixel Group and SRC-B team win the second place and overall third place, respectively.
Second, most methods achieve high PSNR performance (over 40 dB). This indicates most degradations, \eg, glare and haze, are easy to restore.
Only three teams train their models with extra data, and several top-ranked teams apply ensemble strategies (self-ensemble \cite{timofte2016seven}, model ensemble, or both).

\begin{table}[t]
\centering
\caption{Results of MIPI 2022 challenge on UDC image restoration. ‘Runtime’ for per image is tested and averaged across the validation datasets, and the image size is $800\times800$. ‘Params’ denotes the total number of learnable parameters.}
\label{tab:result}
\scalebox{0.6}{
\begin{tabular}{ll|ccc|ccccc}
\toprule
\multicolumn{1}{c}{}                            & \multicolumn{1}{c}{}                            & \multicolumn{3}{|c|}{Metric}                                                               & \multicolumn{1}{l}{}                             &                               &                            &                              &                            \\
\multicolumn{1}{c}{\multirow{-2}{*}{Team Name}} & \multicolumn{1}{c|}{\multirow{-2}{*}{User Name}} & PSNR & SSIM & LPIPS & \multicolumn{1}{l}{\multirow{-2}{*}{Params (M)}} & \multirow{-2}{*}{Runtime (s)} & \multirow{-2}{*}{Platform} & \multirow{-2}{*}{Extra data} & \multirow{-2}{*}{Ensemble} \\
\midrule
USTC\_WXYZ              & YuruiZhu                & $48.48_{(1)}$  & $0.9934_{(1)}$           & $0.0093_{(1)}$        & 16.85             & 0.27      & Nvidia A100                & Yes          & model                      \\
XPixel Group            & JFHu                    & $47.78_{(2)}$  & $0.9913_{(5)}$           & $0.0122_{(6)}$        & 14.06             & 0.16      & Nvidia A6000               & -            & -                          \\
SRC-B                   & xiaozhazha              & $46.92_{(3)}$  & $0.9929_{(2)}$           & $0.0098_{(2)}$        & 23.56             & 0.13      & RTX 3090                   & -            & self-ensemble + model      \\
MIALGO                  & Xhjiang                 & $46.12_{(4)}$  & $0.9892_{(10)}$          & $0.0159_{(13)}$       & 7.93              & 9         & Tesla V100                 & -            & -                          \\
LVGroup\_HFUT           & HuanZheng               & $45.87_{(5)}$  & $0.9920_{(3)}$           & $0.0109_{(4)}$        & 6.47              & 0.92      & GTX 2080Ti                 & -            & -                          \\
GSM                     & Zhenxuan\_Fang          & $45.82_{(6)}$  & $0.9917_{(4)}$           & $0.0106_{(3)}$        & 11.85             & 1.88      & RTX 3090                   & -            & -                          \\
Y2C                     & y2c                     & $45.56_{(7)}$  & $0.9912_{(6)}$           & $0.0129_{(8)}$        &     /              &  /         & GTX 3090                   &              & self-ensemble              \\
VIDAR                   & Null                    & $44.05_{(8)}$  & $0.9908_{(7)}$           & $0.0120_{(5)}$        & 33.8              & 0.47      & RTX 3090Ti                 & -            & model ensemble             \\
IILLab                  & zhuhy                   & $43.45_{(9)}$  & $0.9897_{(9)}$           & $0.0125_{(7)}$        & 82.2              & 2.07      & RTX A6000                  & -            & self-ensemble              \\
jzsherlock              & jzsherlock              & $43.44_{(10)}$ & $0.9899_{(8)}$           & $0.0133_{(9)}$        & 14.82             & 1.03      & Tesla A100                 & -            & self-ensemble              \\
Namecantbenull          & Namecantbenull          & $43.13_{(11)}$ & $0.9872_{(13)}$          & $0.0144_{(10)}$       &    /               &   /       & Tesla A100               &       /       &          /                  \\
MeVision                & Ziqi\_Song              & $42.95_{(12)}$ & $0.9892_{(11)}$          & $0.0150_{(11)}$       &     /              & 0.08      & NVIDIA 3080ti              & Yes          & -                          \\
BIVLab                  & Hao-mk                  & $42.04_{(13)}$ & $0.9873_{(12)}$          & $0.0155_{(12)}$       & 2.98              & 2.52      & Tesla V100                 & -            & self-ensemble              \\
RushRushRush            & stillwaters             & $39.52_{(14)}$ & $0.9820_{(14)}$          & $0.0216_{(14)}$       & 5.9               & 0.63      & RTX 2080Ti                 &      /        &                            \\
JMU-CVLab               & nanashi                 & $37.46_{(15)}$ & $0.9773_{(16)}$          & $0.0370_{(16)}$       & 2                 & 0.48      & Tesla P100                 & -            & -                          \\
eye3                    & SummerinSummer          & $36.70_{(16)}$ & $0.9783_{(15)}$          & $0.0326_{(15)}$       & 26.13             & 1.07      & Tesla V100                 & -            & -                          \\
FMS Lab                 & hrishikeshps94          & $35.77_{(17)}$ & $0.9719_{(17)}$          & $0.0458_{(18)}$       & 4.40              & 0.04      & Tesla V100                 & -            & -                          \\
EDLC2004                & jiangbiao               & $35.50_{(18)}$ & $0.9616_{(18)}$          & $0.0453_{(17)}$       & 31                & 0.99      & RTX 2080Ti                 & Yes          & -                          \\
SAU\_LCFC               & chm                     & $32.75_{(19)}$ & $0.9591_{(19)}$          & $0.0566_{(19)}$       & 0.98              & 0.92      & Tesla V100                 &     -         &               -            \\
\bottomrule
\end{tabular}}
\end{table}

\section{Challenge Methods and Teams}
\label{sec:methods}

\begin{figure}[t]
    \centering
     \includegraphics[width=0.8\textwidth]{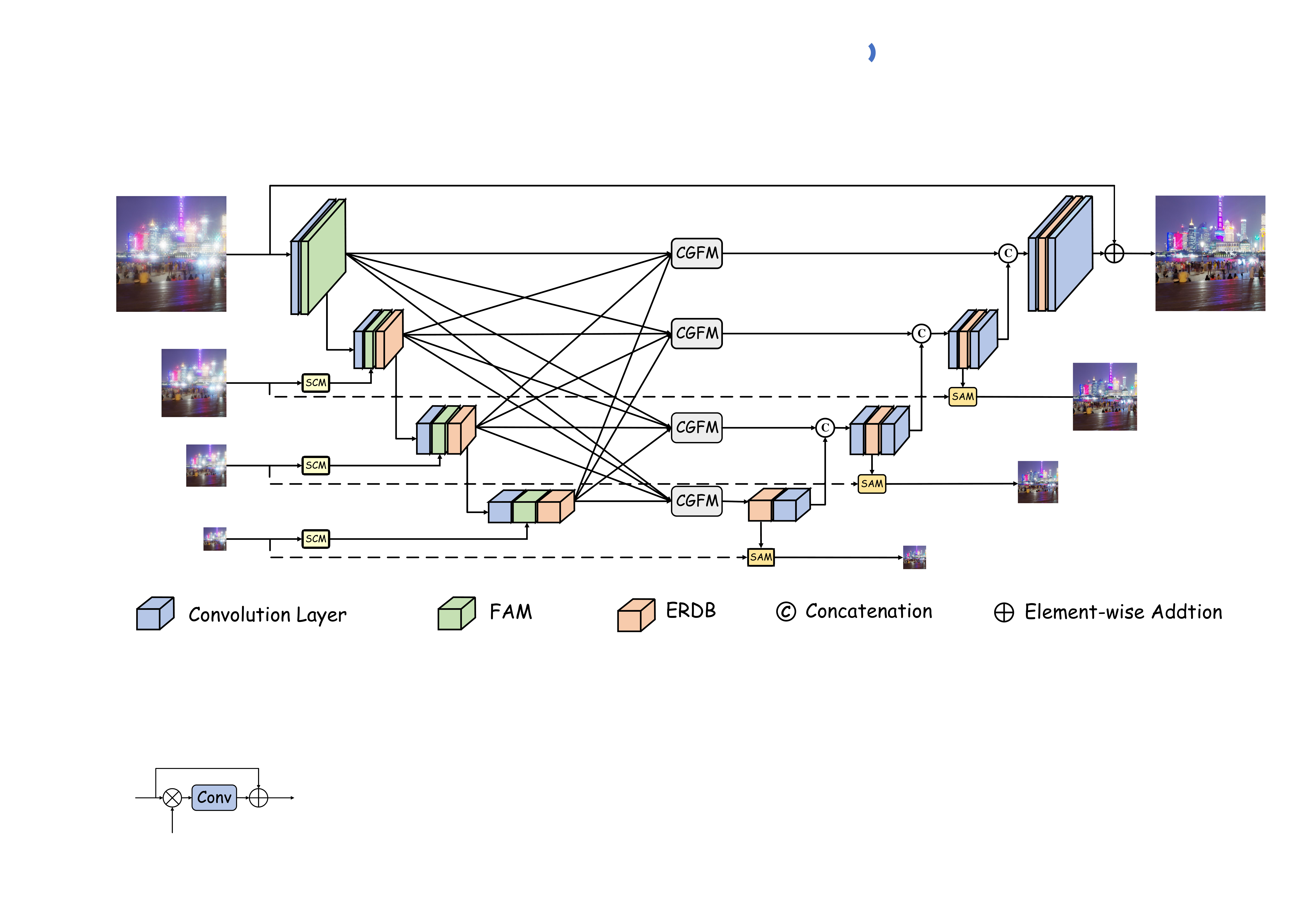}
     \includegraphics[width=0.8\textwidth]{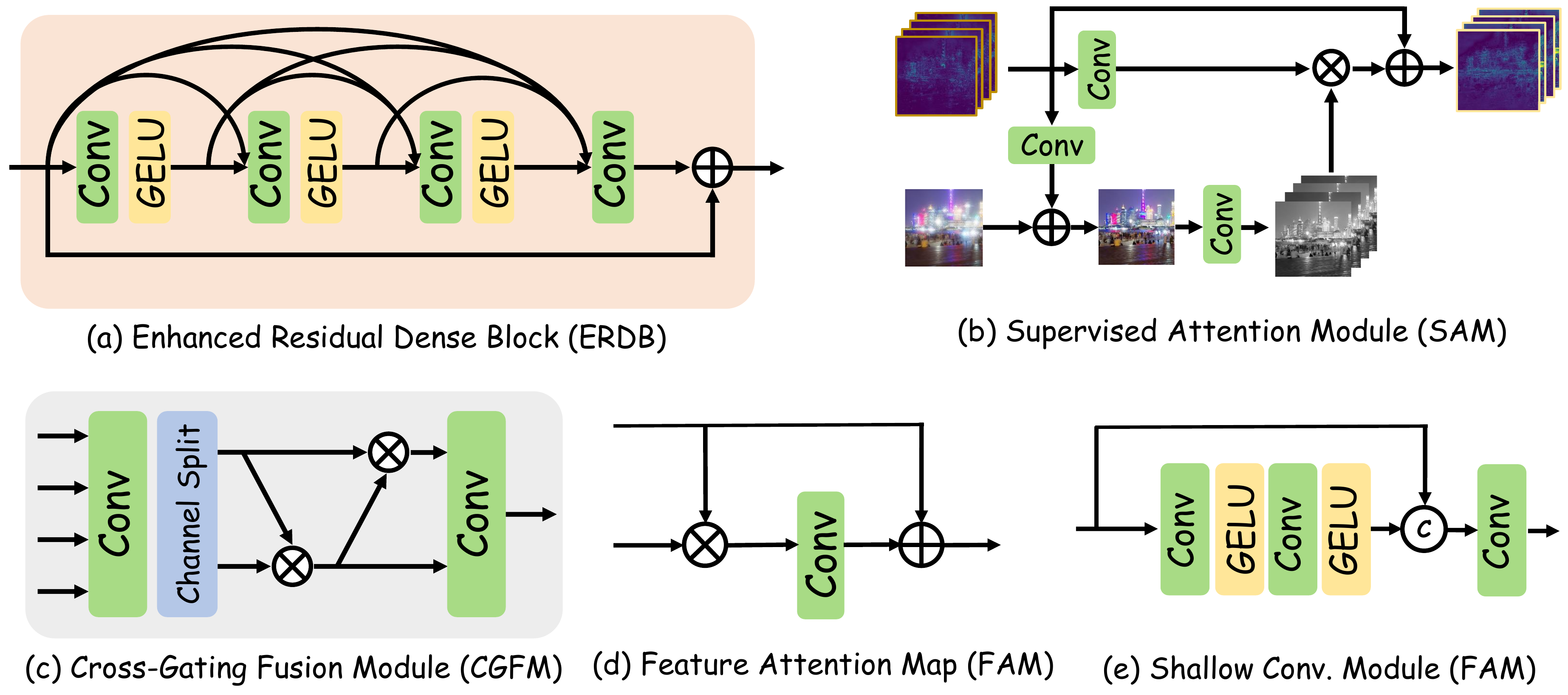}
    \caption{The network architecture proposed by USTC\_WXYZ Team.}
    \label{fig:USTC_WXYZ}
    \vspace{-0.45cm}
\end{figure}

\paragraph{\bf USTC\_WXYZ Team.}
Inspired by \cite{cho2021rethinking,huang2017densely,zamir2021multi}, this team designs an enhanced multi-inputs multi-outputs network, which mainly consists of dense residual blocks and the cross-scale gating fusion modules.
The overall architecture of the network is depicted in Figure~\ref{fig:USTC_WXYZ}.
The training phase could be divided into three stages:
i) Adopt the Adam optimizer with a batch size of $3$ and the patch size of $256\times256$. The initial learning rate is $2\times 10^{-4}$ and is adjusted with the Cosine Annealing scheme, including $1000$ epochs in total.
ii) Adopt the Adam optimizer with a batch size of $1$ and the patch size of $512\times512$. The initial learning rate is $2\times 10^{-5}$ and is adjusted with the Cosine Annealing scheme, including $300$ epochs in total.
iii) Adopt the Adam optimizer with a batch size of $2$ and the patch size of $800\times800$. The initial learning rate is $8\times 10^{-6}$ and is adjusted with the Cosine Annealing scheme, including $150$ epochs in total.
During inference, the team adopts model ensemble strategy averaging the parameters of multiple models trained with different hyperparameters, which brings around 0.09 dB increase on PSNR.

\begin{figure}[t]
    \centering
     \includegraphics[width=0.8\textwidth]{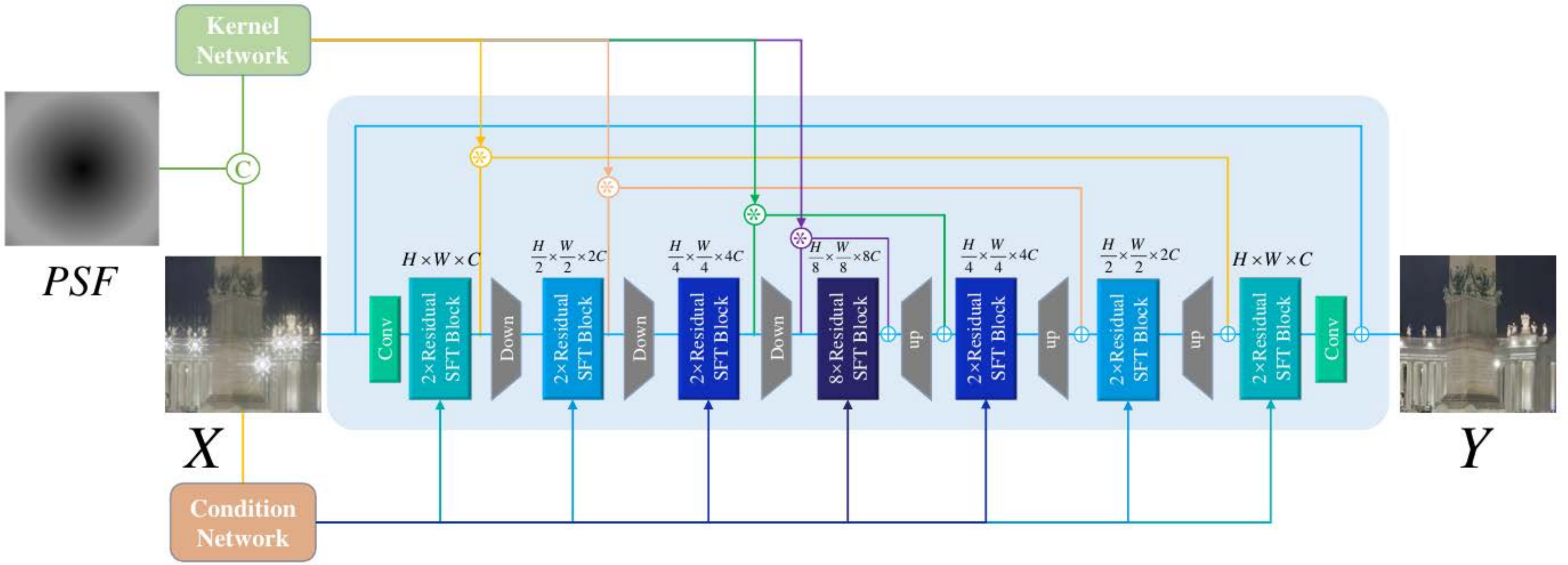}
    \caption{The network architecture proposed by XPixel Group.}
    \label{fig:xpixel}
    \vspace{-0.5cm}
\end{figure}

\paragraph{\bf XPixel Group.}
This team designs a UNet-like structure (see Figure~\ref{fig:xpixel}), making full use of the hierarchical multi-scale information from low-level features to high-level features. To ease the training procedure and facilitate the information flow, several residual blocks are utilized in the base network, and the dynamic kernels in the skip connection bring better flexibility. Since UDC images have different holistic brightness and contrast information, the team applies the condition network with spatial feature transform (SFT) to deal with input images with location-specific and image-specific operations (mentioned in HDRUNet \cite{chen2021hdrunet}), which could provide spatially variant manipulations. In addition, they incorporate the information of the point spread function (PSF) provided in DISCNet \cite{feng2021removing}, which has demonstrated its effectiveness through extensive experiments on both synthetic and real UDC data.
For training, the authors randomly cropped $256\times256$ patches from the training images as inputs. The mini-batch size is set to $32$ and the whole network is trained for $6\times 10^5$ iterations. The learning rate is initialized as $2\times 10^{-4}$, decayed with a CosineAnnealing schedule, and restarted at $[50K, 150K, 300K, 450K]$ iterations.

\begin{figure}[t]
    \centering
     \includegraphics[width=0.8\textwidth]{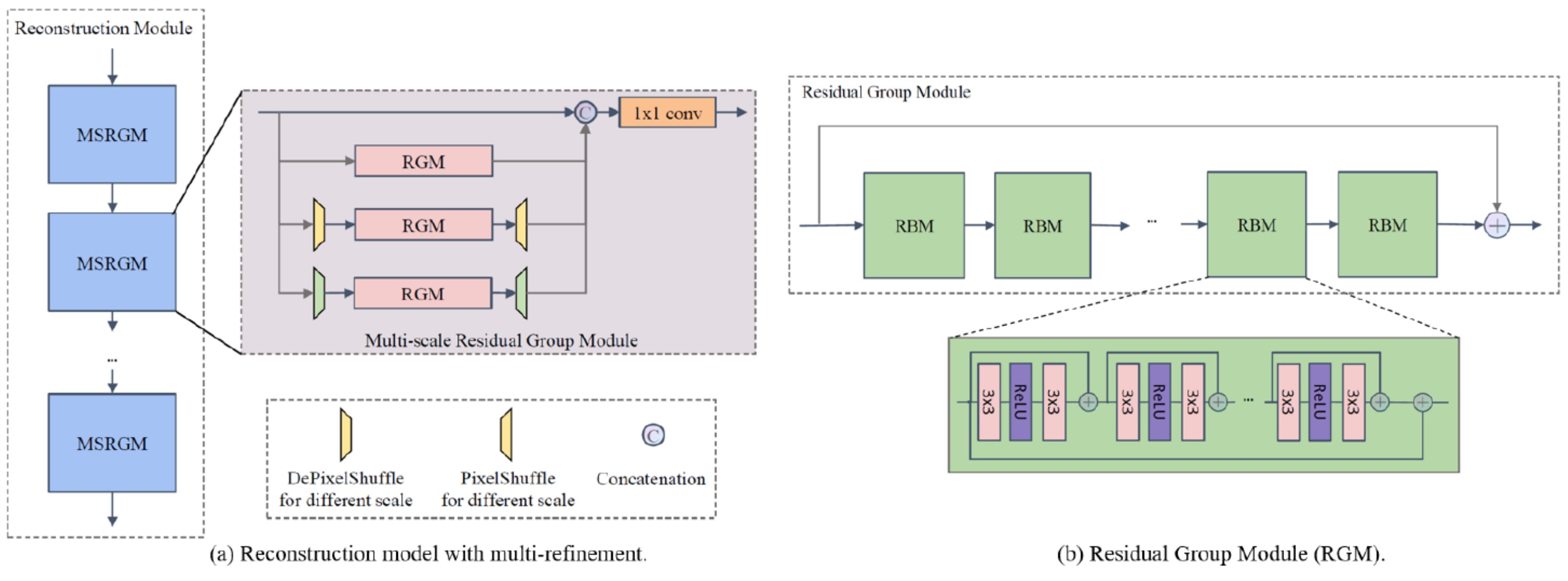}
    \caption{The network architecture proposed by SRC-B Team.}
    \label{fig:src_b}
    \vspace{-0.5cm}
\end{figure}

\paragraph{\bf SRC-B Team.}
This team proposes Multi-Refinement Network (MRNet) for Image Restoration on Under-Display Camera, as shown in Figure~\ref{fig:src_b}.
In this challenge, the authors modify and improve the MRNet \cite{abuolaim2021ntire}, which is mainly composed of 3 modules: shallow Feature Extraction, reconstruction module, and output module. The shallow Feature Extraction and output module only use one convolution layer.
Multi-refinement is the main idea of the reconstruction module, which includes $N$ Multi-scale Residual Group Module (MSRGM) and progressively refines the features. MSRGM is also used to fuse information from three scales to improve the representation ability and robustness of the model, where each scale is composed of several residual group modules (RGM). RGM contains $G$ residual block module (RBM). In addition, the authors propose to remove Channel Attention (CA) module in RBM \cite{zhang2018image}, since it brings limited improvement but increases the inference time.

\begin{figure}[t]
    \centering
     \includegraphics[width=0.8\textwidth]{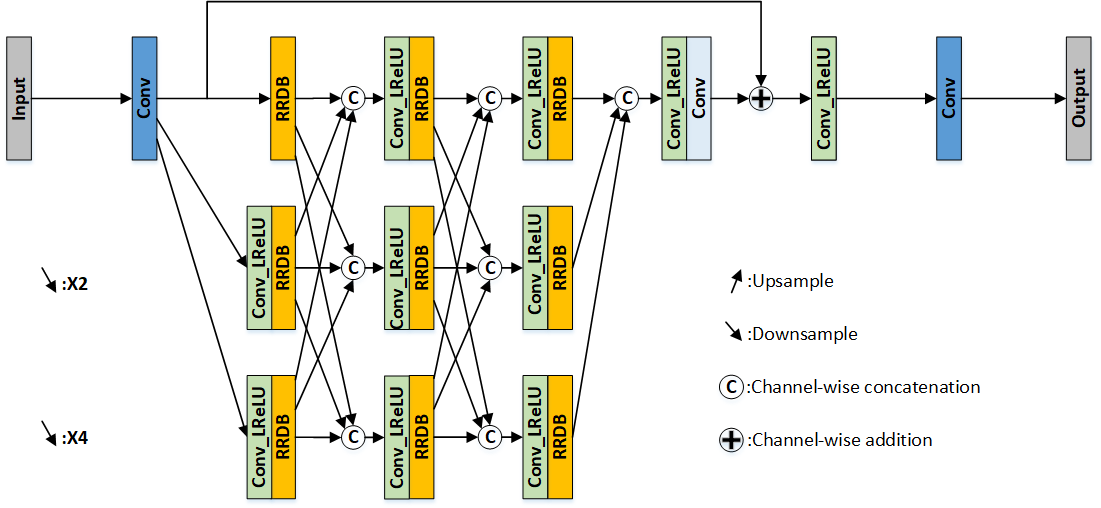}
     \begin{subfigure}[b]{0.4\textwidth}
      \includegraphics[width=0.8\textwidth]{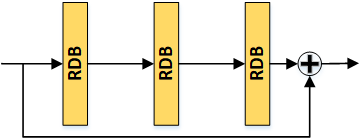}
      \caption{(a) RRDB}
     \end{subfigure}
     \begin{subfigure}[b]{0.4\textwidth}
      \includegraphics[width=\textwidth]{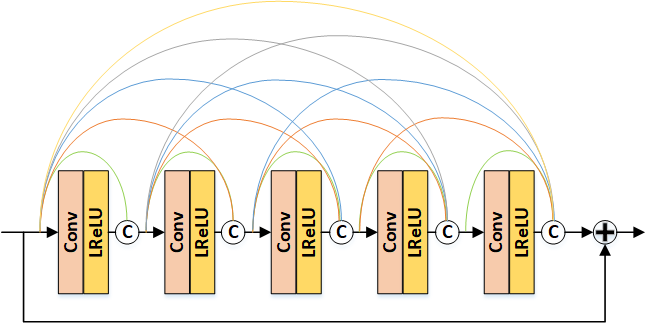}
      \caption{(b) RDB}
     \end{subfigure}
    \caption{The network architecture proposed by MIALGO Team.}
    \label{fig:mialgo}
    \vspace{-0.5cm}
\end{figure}

\paragraph{\bf MIALGO Team.}
As shown in Figure~\ref{fig:mialgo}, this team addresses the UDC image restoration problem with an Residual Dense Network (RDN) \cite{zhang2018residual} as the backbone. The authors also adopt the strategy of multi-resolution feature fusion in HRNet \cite{sun2019deep} to improve the performance.
In order to facilitate the recovery of high dynamic range images, the authors generate 6 non-HDR images with different exposure ranges from the given HDR images, where the value ranges of each split are: [0, 0.5], [0, 2], [0, 8], [0, 32], [0, 128], [0, 500], and normalize them and then stack into 18 channels of data for training. According to the dynamic range of the data, 2000 images in the training datasets (the remaining 16 are used as validation sets) are divided into two datasets:
i) Easy sample dataset: 1318 images whose intensities lie in [0, 32].
ii) Hard sample dataset: easy samples plus doubled remaining hard samples.
They use two datasets to train two models, each of which is first trained with 500k iterations. Then they fine-tune the easy sample model and the hard sample model by 200K iterations. In the validation/test phase, the 18-channel data is merged into 3-channel HDR images and fed to the network.

\begin{figure}[t]
    \centering
     \includegraphics[width=0.8\textwidth]{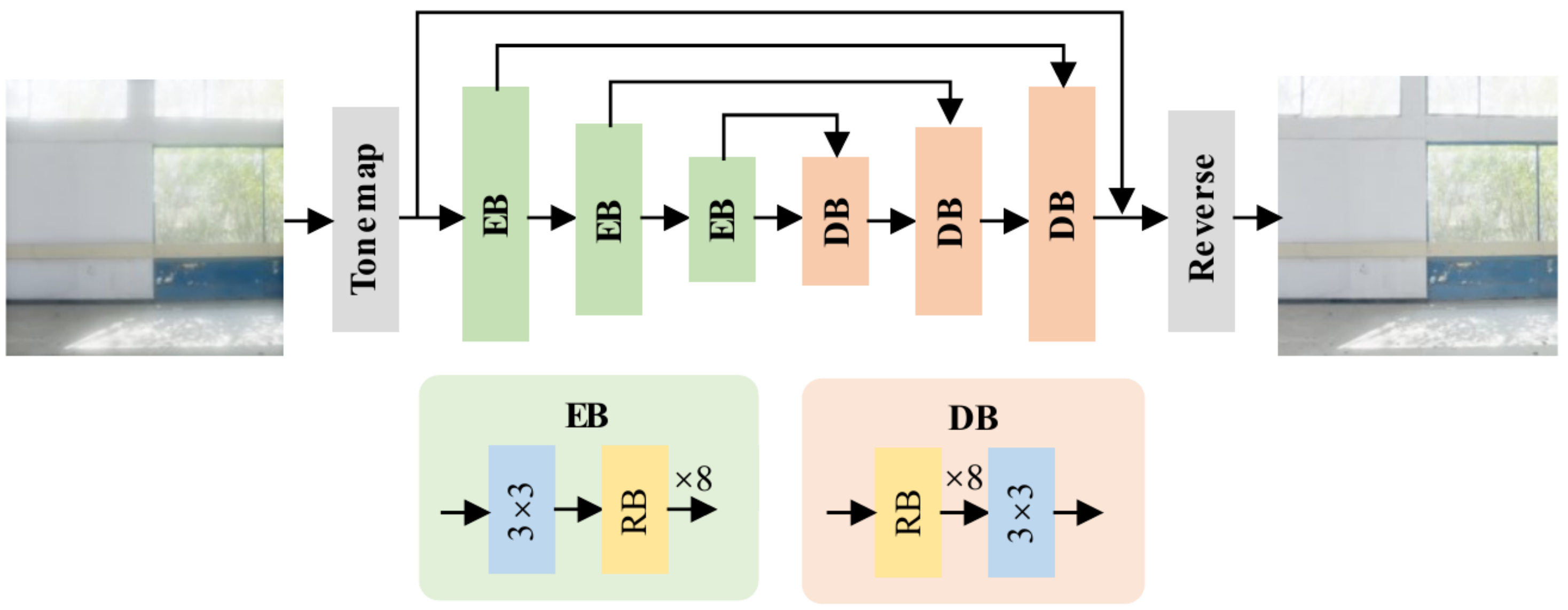}
    \caption{The network architecture proposed by LVGroup\_HFUT Team.}
    \label{fig:lvgroup}
    \vspace{-0.5cm}
\end{figure}

\paragraph{\bf LVGroup\_HFUT Team.}
Considering the specificity of the UDC image restoration task (mobile deployment), this team designs a lightweight and real-time model for UDC image restoration using a simple UNet architecture with magic modifications as shown in Figure~\ref{fig:lvgroup}. Specifically, they combine the full-resolution network FRC-Net \cite{zhang2022frc} and the classical UNet \cite{ronneberger2015u} to construct the model, which presents two following advantages: 1) directly stacking the residual blocks at the original resolution to ensure that the model learns sufficient spatial structure information, and 2) stacking the residual blocks at the downsampled resolution to ensure that the model learns sufficient semantic-level information.

During training, they first perform tone mapping, and then perform a series of data augmentation sequentially, including: 1) random crop to $384\times384$; 2) vertical flip with probability $0.5$; 3): horizontal flip with probability $0.5$. They train the model for $3000$ epochs on provided training dataset with an initial learning rate $1e-4$ and batch size 4.

\begin{figure}[t]
    \centering
     \includegraphics[width=0.8\textwidth]{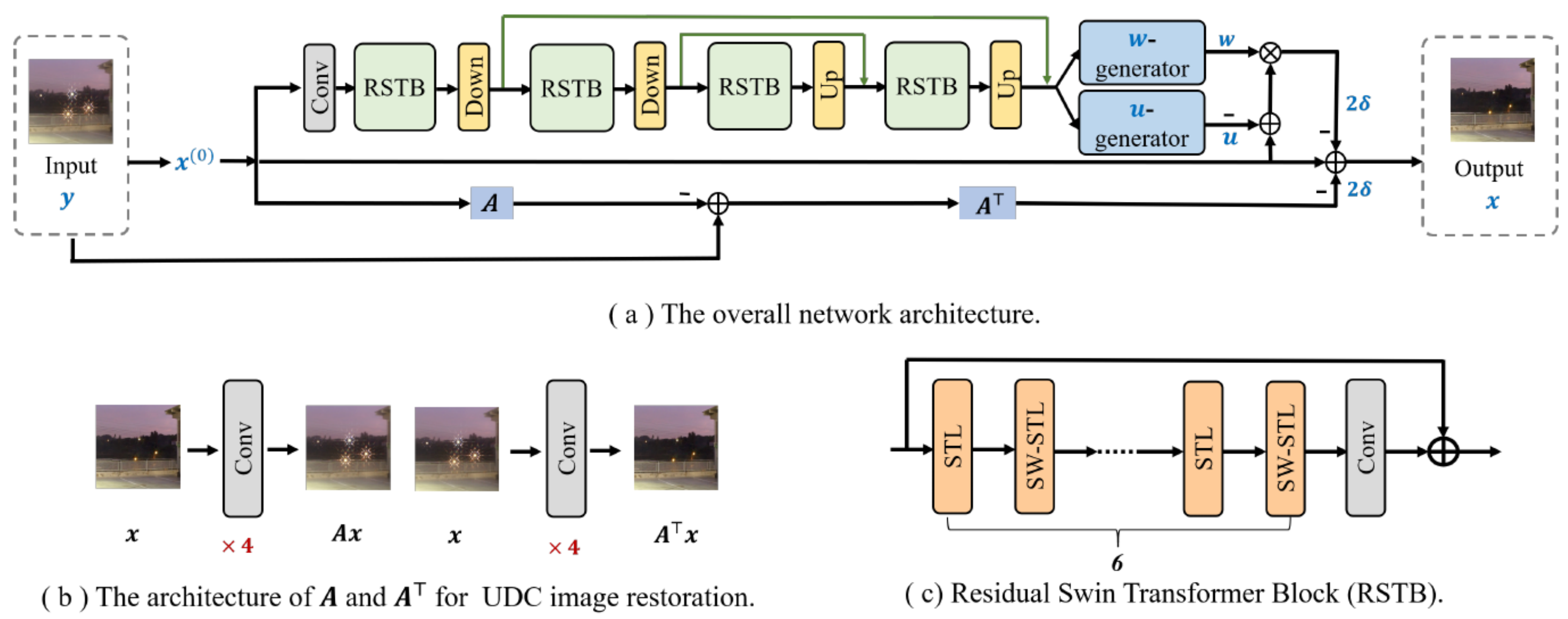}
    \caption{The network architecture proposed by GSM Team.}
    \label{fig:gsm}
    \vspace{-0.5cm}
\end{figure}

\paragraph{\bf GSM Team.}
This team reformulates UDC image restoration as a Maximum  Posteriori (MAP) estimation problem with the learned Gaussian Scale Mixture (GSM) models. Specifically, the $x$ can be solved by
\begin{equation}
\label{eqn:gsm}
    \bm{x}^{(t+1)}= \bm{x}^{(t)}-2\delta\{\textbf{A}^T(\textbf{A}\bm{x}^{(t)}-\bm{y})+\bm{w}^{(t)}(\bm{x}^{(t)}-\bm{u}^{(t)})\},
\end{equation}
where $\textbf{A}^T$ denotes the transposed version of $\textbf{A}$, $\delta$ denotes the step size, $t$ denotes the $t^{th}$ iteration for iteratively optimizing $\bm{x}$, $\bm{w}^{(t)}$ denotes the regularization parameters, and $\bm{u}^{(t)}$ denotes the mean of the GSM model.
In \cite{huang2021deep}, a UNet was used to estimated $\bm{w}^{(t)}$ and $\bm{u}^{(t)}$ and two sub-networks with 4 Resblocks and 2 Conv layers were used to learn $\textbf{A}$ and $\textbf{A}^T$, respectively. For UDC image restoration, they develop a network based on Swin Transformer to learn the GSM prior (\ie, $\bm{w}^{(t)}$ and $\bm{u}^{(t)}$) and use 4 Conv layers, respectively. As shown in Figure~\ref{fig:gsm}, the team constructs an end-to-end network for UDC image restoration. The transformer-based GSM prior network contains an embedding layer, $4$ Residual Swin Transformer Blocks (RSTB) \cite{liang2021swinir}, two downsampling layers, two upsampling layers, a $w$-generator, and an $u$-generator. The embedding layer, the $w$-generator, and the $u$-generator are a $3\times3$ Conv layer. The RSTB \cite{liang2021swinir} contains 6 Swin Transformer Layers, a Conv layer, and a skip connection. The features of the first two RSTBs are reused by two skip connections, respectively. For reducing the computational complexity, they implement the united framework Eq.~\ref{eqn:gsm} with only one iteration and use the input $\bm{y}$ as the initial value $\bm{x}^{(0)}$.

The loss function $L$ is defined as
\begin{equation}
    \begin{aligned}
        & L_1 = \sqrt{\|\bm{x}-\hat{\bm{x}}\|^2+\epsilon},\\
        & L_2 = \sqrt{\|\Delta(\bm{x})-\Delta(\hat{\bm{x}})\|^2+\epsilon},\\
        & L_3 = \|\mathcal{FFT}(\bm{x})-\mathcal{FFT}(\hat{\bm{x}})\|_1,\\
        & L = L_1+0.05\times L_2 + 0.01\times L_3,
    \end{aligned}
\end{equation}
where $\bm{x}$ and $\bm{\hat{x}}$ denote the label and the output of the network, $\Delta$ represents the Laplacian operator, and $\mathcal{FFT}$ is the FFT operation.

\begin{figure}[t]
    \centering
     \includegraphics[width=0.8\textwidth]{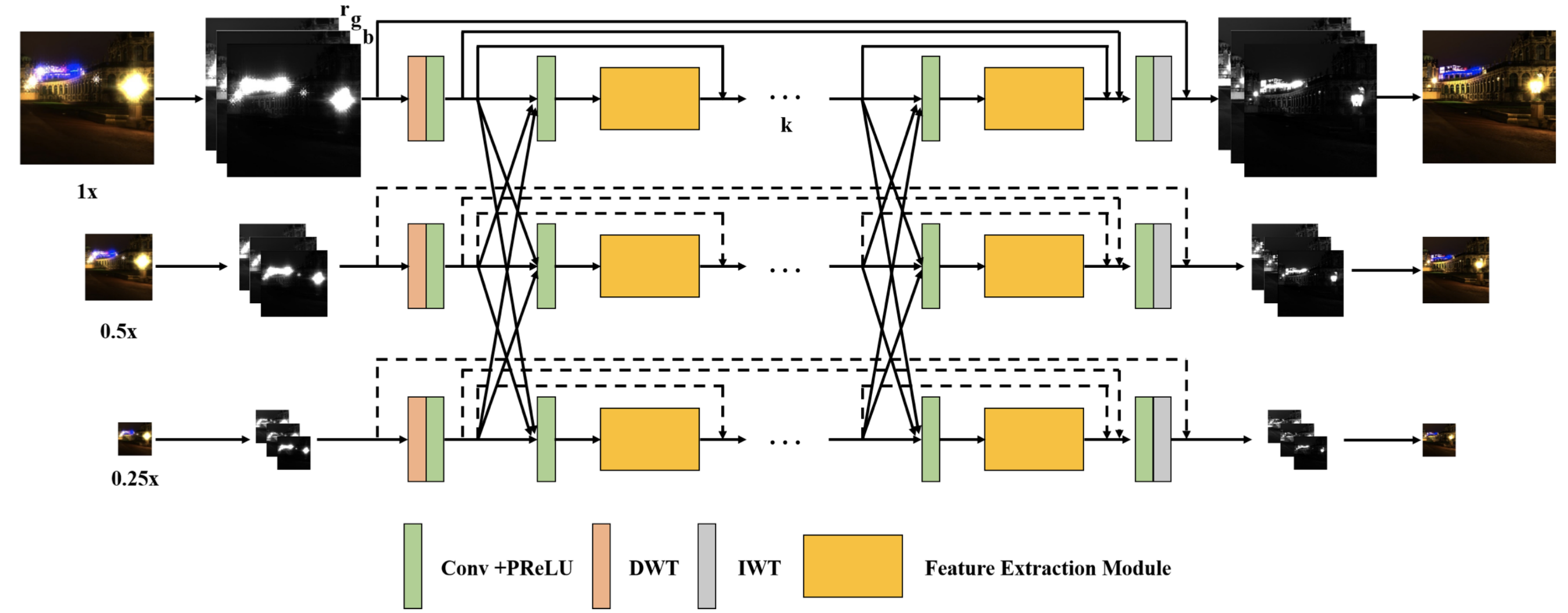}
    \caption{The network architecture proposed by Y2C Team.}
    \label{fig:y2c}
    \vspace{-0.5cm}
\end{figure}

\paragraph{\bf Y2C Team.}
This team takes model \cite{cho2021rethinking,zamir2022learning} as a backbone network and introduces the multi-scale design. Besides, they reconstruct each color channel by a complete branch instead of processing them together. The overall architecture of the Multi-Scale and Separable Channels reconstruction Network (MSSCN) for UDC image restoration is shown in Figure~\ref{fig:y2c}. The network takes the multi-scale degraded images (1x, 0.5x, 0.25x) as inputs, and the initial feature maps of each color channel in each scale are extracted respectively. Before extracting feature maps, they use the discrete wavelet transform (DWT) \cite{liu2018multi} to reduce the resolution in each scale and improve the reconstruction efficiency. In the fusion module, this team fuses the feature maps in different scales in the same spatial resolution, concatenates the feature maps from each color channel and each scale, and processes them using a depth-wise convolution. In the feature extraction module, any efficient feature extraction block can be used to extract feature maps. Residual group (RG) with multiple residual channel attention blocks (RCAB) \cite{zhang2018image} is used as the feature extraction module. To improve the learning capacity of recovering images, they repeat the feature fusion module and feature extraction module $k$ times. Then the reconstructed images are generated by the inverse discrete wavelet transform (IWT).

\begin{figure}[t]
    \centering
     \includegraphics[width=0.8\textwidth]{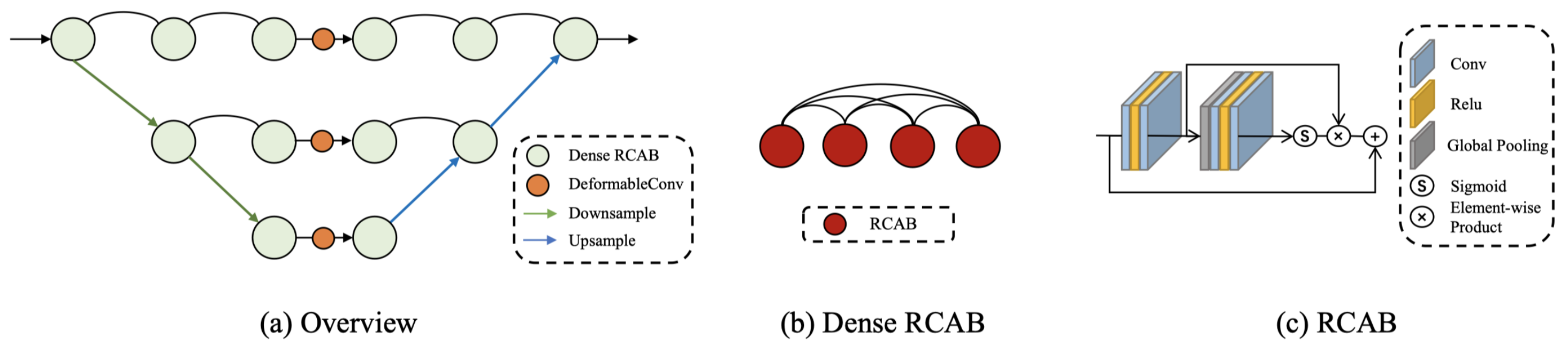}
    \caption{The network architecture proposed by VIDAR Team.}
    \label{fig:vidar}
    \vspace{-0.5cm}
\end{figure}

\paragraph{\bf VIDAR Team.}
As shown in Figure~\ref{fig:vidar}, this team designs a dense U-Net for under-display camera image restoration.
The architecture aggregates and fuses feature information with Dense RCAB \cite{zhang2018image}.
During training, batch size is set to 4 with the patch of $384 \times 384$ and the optimizer is ADAM \cite{kingma2014adam} by setting $\beta_1=0.9$, $\beta_2=0.999$. The initial learning rate is $1 \times 10^{-4}$. They use Charbonnier loss to train models and stop training when no notable decay of training loss is observed.

\begin{figure}[t]
    \centering
     \includegraphics[width=0.8\textwidth]{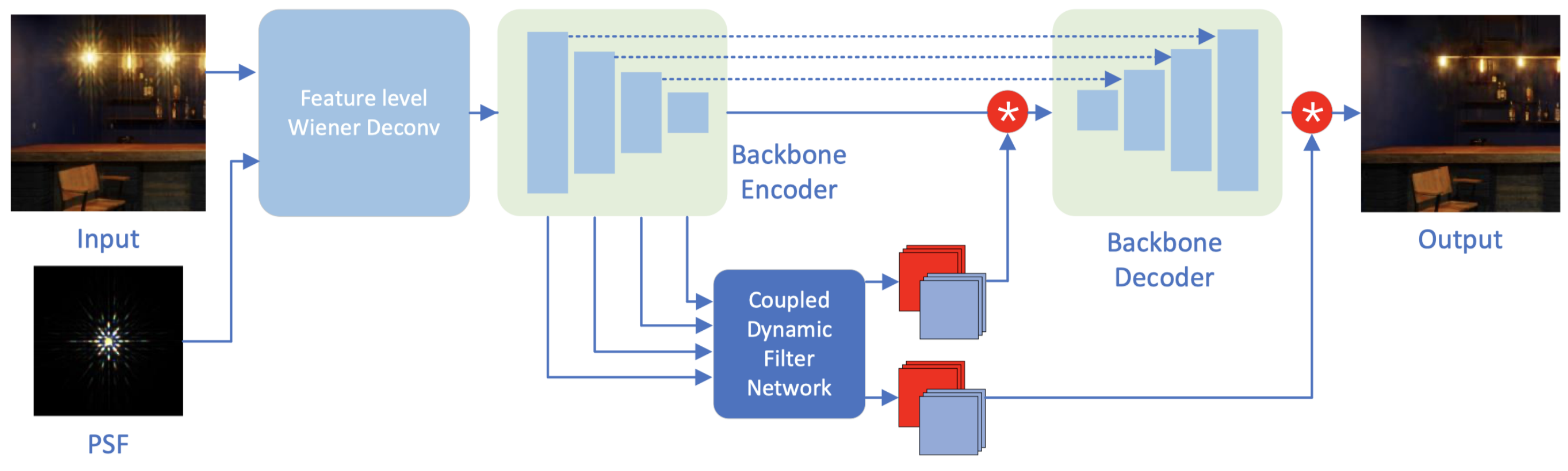}
    \caption{The network architecture proposed by IILLab Team.}
    \label{fig:iillab}
    \vspace{-0.5cm}
\end{figure}

\paragraph{\bf IILLab Team.}
This team presents a Wiener Guided Coupled Dynamic Filter Network, which takes advantage of both DWDN \cite{dong2020deep} and BPN \cite{xia2020basis}. As shown in Figure~\ref{fig:iillab}, on top of the U-net-based backbone, they first used feature level wiener for global restoration with the estimated degradation kernel, then used the coupled dynamic filter network for local restoration. The network obtains the global and local information from the feature level wiener deconvolution and the coupled dynamic filter network, and thus achieves a better restoration result since the UDC degradation was modeled with higher accuracy.
Wiener Guided Coupled Dynamic Filter Network was jointly trained under the supervision of L1 loss and perceptual loss \cite{johnson2016perceptual}.

\begin{figure}[t]
    \centering
     \includegraphics[width=0.8\textwidth]{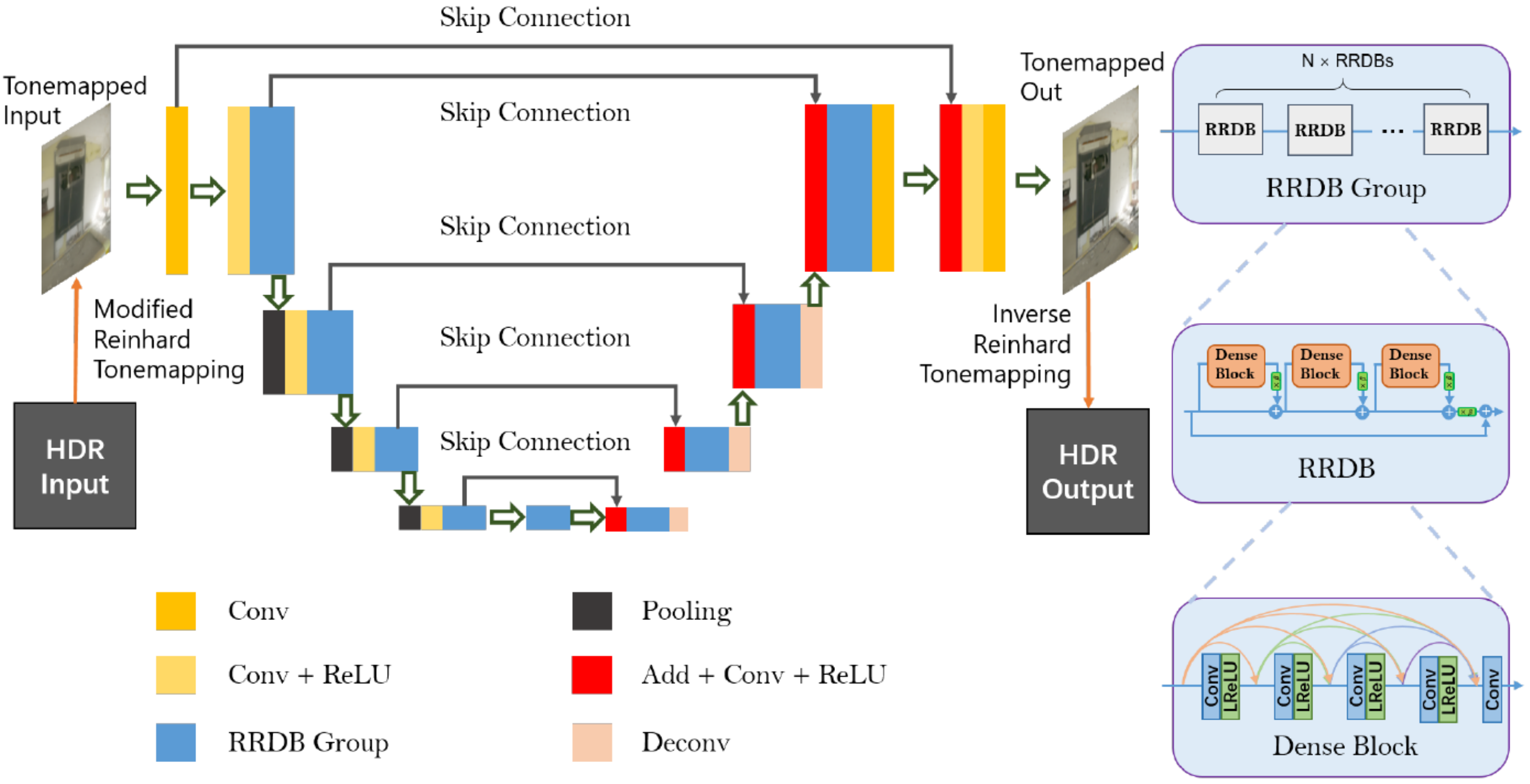}
    \caption{The network architecture proposed by jzsherlock Team.}
    \label{fig:jzsherlock}
    \vspace{-0.5cm}
\end{figure}

\paragraph{\bf jzsherlock Team.}
This team proposes the U-RRDBNet for the task of UDC image restoration. UDC images suffer from the strong glare caused by the optical structure of OLED display, which usually occurs and degrades a large area around the light source (\eg street lamp, sun) in the view.
Therefore a larger receptive field is required to deal with this task.
The proposed architecture and processing pipeline are illustrated in Figure~\ref{fig:jzsherlock}. Though the original U-net and its variants have achieved promising results on segmentation tasks, the capability of representation learning of the U-net is still limited for dense prediction in low-level tasks. So the authors apply the Residual-in-Residual Dense Block (RRDB) \cite{wang2018esrgan} to substitute simple Conv layers in U-net after each down/up-sampling operation to increase the network capability.
They first perform the modified Reinhard tone mapping, and then feed the tone-mapped image into the end-to-end U-RRDBNet for restoration, and produce the image in the tone-mapped domain with artifacts removed.
The output of the network is then transformed back to the HDR domain using the inverse of Reinhard tone mapping. The model is first trained using $L_1$ loss and perceptual loss \cite{johnson2016perceptual} in the tone-mapped domain, and then fine-tuned with MSE loss for higher PSNR performance.

\begin{figure}[t]
    \centering
     \includegraphics[width=0.8\textwidth]{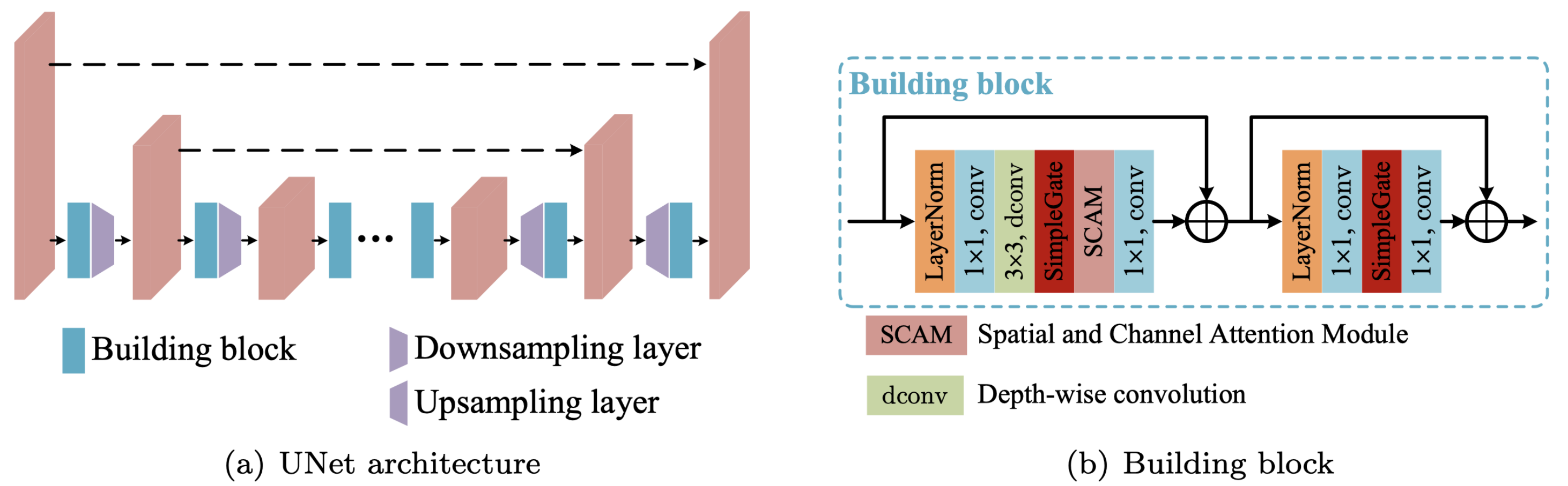}
    \caption{The network architecture proposed by Namecantbenull Team.}
    \label{fig:namecantbenull}
    \vspace{-0.5cm}
\end{figure}

\paragraph{\bf Namecantbenull Team.}
This team designs a deep learning model for the UDC image restoration based on the U-shaped network. In this network, they incorporate the convolution operation and attention mechanism into the transformer block architecture, which has been proved effective in image restoration tasks in \cite{chen2022simple}. Since there is no need to calculate the self-attention coefficient matrices, the memory cost and computation complexity can be reduced significantly. The overall network structure is shown in Figure~\ref{fig:namecantbenull}. Specifically, they substitute the simplified channel attention module with the spatial and channel attention module \cite{chen2017sca} to additionally consider the correlation of the spatial pixels. They train the model with $L$1 and perceptual loss on the tone-mapped domain.

\begin{figure}[t]
    \centering
     \includegraphics[width=0.8\textwidth]{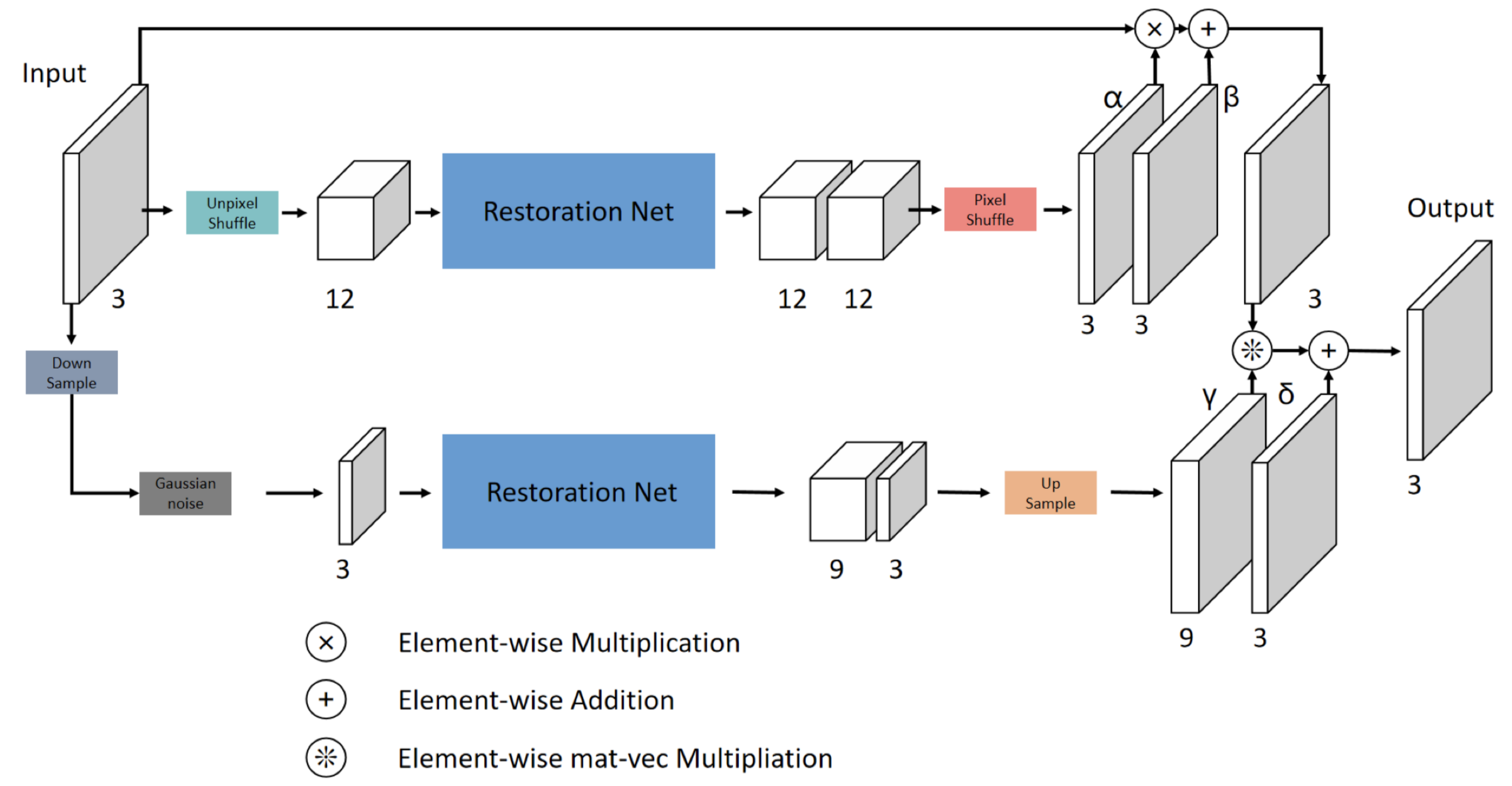}
     \includegraphics[width=0.8\textwidth]{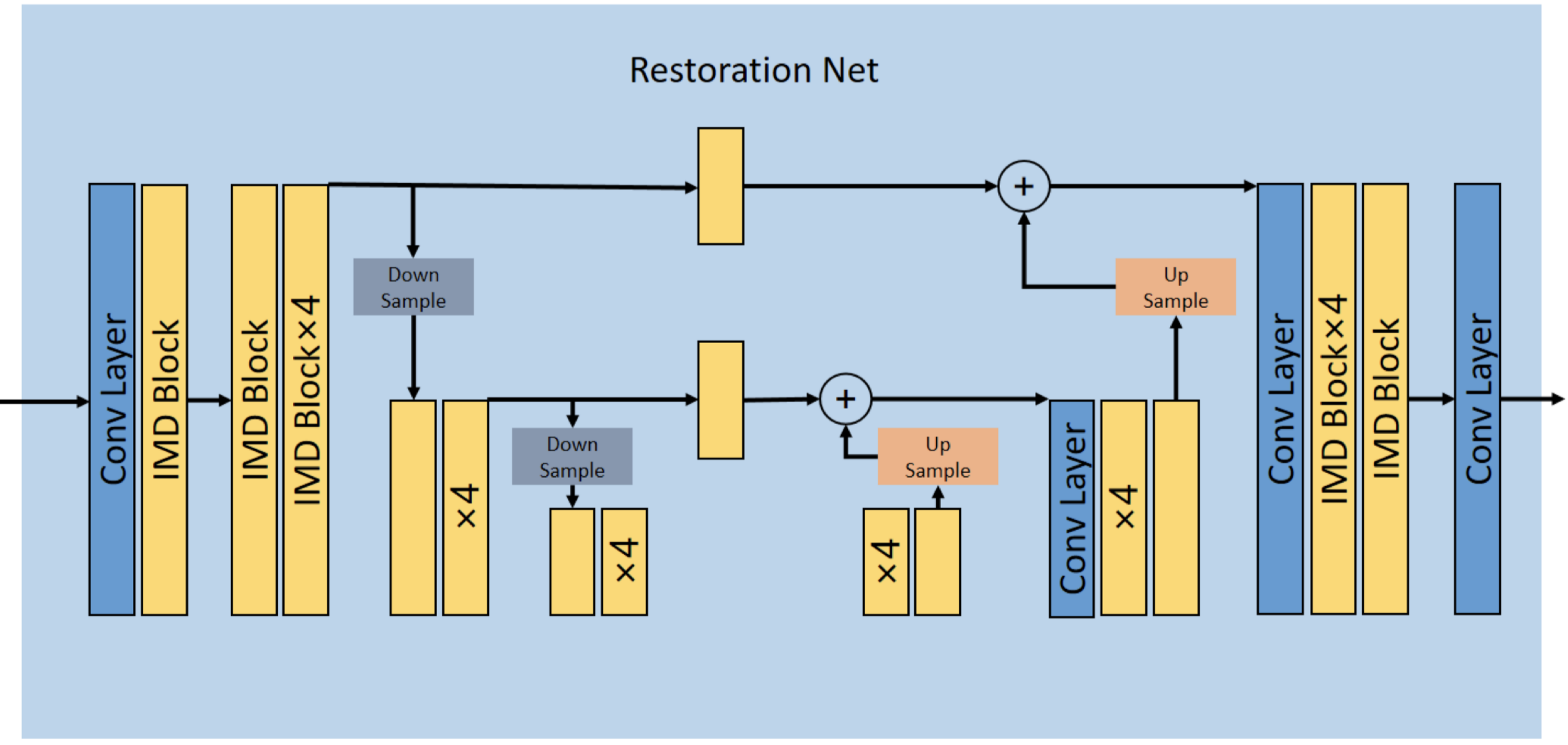}
    \caption{The network architecture proposed by MeVision Team.}
    \label{fig:mevision}
    \vspace{-0.5cm}
\end{figure}

\paragraph{\bf MeVision Team.}
This team follows \cite{koh2022bnudc} and presents a two-branch network to restore UDC images. As shown in Figure~\ref{fig:mevision}, the input images are processed and fed to the restoration network. The original resolution image is processed in one branch while the blurred image with noise is processed in another one. Finally, the two branches are connected in series by the affine transformation connection.
Then, they follow \cite{mao2016image} and modify the high-frequency reconstruction branch with an encoder-decoder structure to reduce parameters. And they converted the smoothed diluted residual block into IMD block \cite{hui2019lightweight} so that the signals can propagate directly from skip connections to the bottom layers.

\begin{figure}[t]
    \centering
     \includegraphics[width=0.8\textwidth]{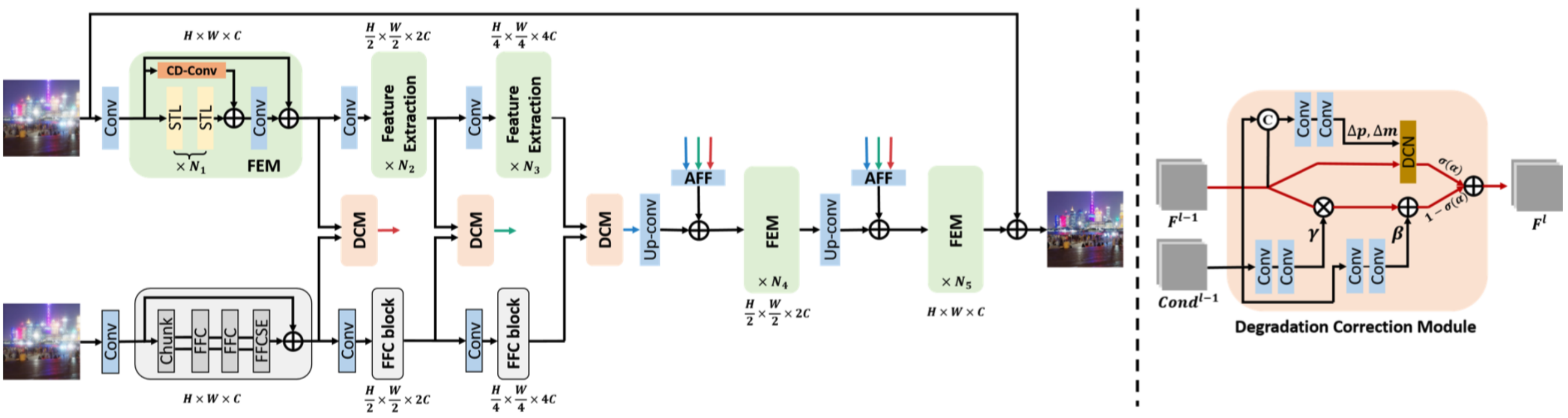}
    \caption{The network architecture proposed by BIVlab Team.}
    \label{fig:bivlab}
    \vspace{-0.7cm}
\end{figure}

\paragraph{\bf BIVLab Team.}
This team develops a Self-Guided Coarse-to-Fine Network (SG-CFN) to progressively reconstruct the degraded under-display camera images. As shown in Figure~\ref{fig:bivlab}, the proposed SG-CFN consists of two branches, the restoration branch and the condition branch. The restoration branch is constructed by Feature Extraction Module (FEM) based on improved RSTBs \cite{liang2021swinir}, which incorporates the paralleled Central Difference Convolution (CDC) with Swin Transformer Layer (STL) to extract rich features at multiple scales. The condition branch is constructed by Fast Fourier Convolution (FFC) blocks \cite{chi2020fast} that endowed with global receptive field to model the distribution of the large-scale point-spread function (PSF), which is indeed the culprit of the degradation. Furthermore, the multi-scale representations extracted from the restoration branch are processed via the Degradation Correction Module (DCM) to restore clean features, guided by the corresponding condition information. To fully exploit the restored multi-scale clean features, they enable the asymmetric feature fusion inspired by \cite{cho2021rethinking} to facilitate the flexible information flow.
Images captured through under-display camera typically suffer from diffraction degradation caused by large-scale PSF, resulting in blurred and detail attenuated images. Thereby, (1) incorporating CDC into the restoration branch can greatly help to avoid over-smooth feature extraction and enrich the representation capability, (2) constructing the condition branch with FFC blocks endow the global receptive field to capture the degradation information, and (3) correcting the degraded features from both local adaptation and global modulation makes the process of restoration more effective.
During the testing phase, they adopt self-ensemble strategy and it brings a 0.63 dB performance gain on PSNR.

\paragraph{\bf RushRushRush Team.}
This team reproduced the MIRNetV2 \cite{Zamir2022MIRNetv2} on the UDC dataset.
The network utilizes both spatially-precise high-resolution representations and contextual information from the low-resolution representations. The multi-scale residual block contains: (i) parallel multi-resolution convolution streams to extract multi-scale features, (ii) information exchange across streams, (iii) non-local attention mechanism to capture contextual information, and (iv) attention-based multi-scale feature aggregation. The approach learns an enriched set of features that combines contextual information from multiple scales, while simultaneously preserving the high-resolution spatial details.

\begin{figure}[t]
    \centering
     \includegraphics[width=0.8\textwidth]{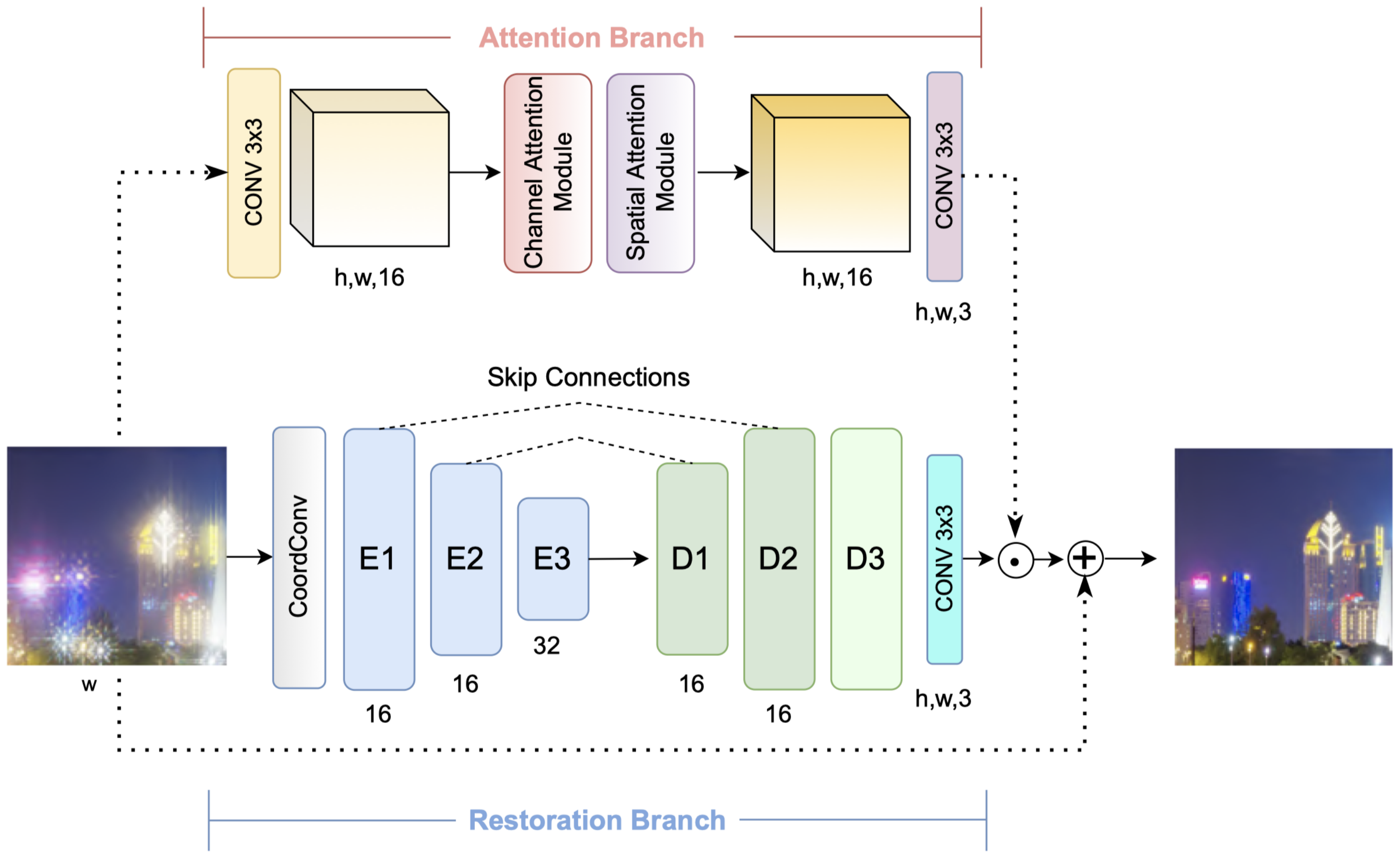}
    \caption{The network architecture proposed by JMU-CVLab Team.}
    \label{fig:jmu_cvlab}
    \vspace{-0.7cm}
\end{figure}

\paragraph{\bf JMU-CVLab Team.}
This team proposes a dual-branch lightweight neural network.
Inspired by recent camera modeling approaches \cite{conde2022model}, they design a dual-branch model, depending on the tradeoff between resources and performance.
As shown in Figure~\ref{fig:jmu_cvlab}, the method combines ideas from deblurring \cite{nah2021ntire} and HDR \cite{liu2020single} networks, and attention methods \cite{liu2018intriguing,woo2018cbam}. This does not rely on metadata or extra information like the PSF.
The main image restoration branch with a Dense Residual UNet architecture \cite{zhang2018residual,ronneberger2015u}. They use an initial CoordConv \cite{liu2018intriguing} layer to encode positional information. Three Encoder blocks, each formed by 2 dense residual layers (DRL) \cite{zhang2018residual} followed by the corresponding downsampling pooling. The decoder blocks D$_1$ and D$_2$ consist of a bilinear upsampling layer and 2 DRL The decoder block D$_3$ does not have an upsampling layer, consists of 2 DRL and a series of convolutions with 7, 5, and 1 kernel sizes, the final convolution (cyan color) produces the residual using a tanh activation.
Similar to \cite{liu2020single}, the additional attention branch aims to generate an attention map (per channel) to control the hallucination in overexposed areas. The attention map is generated after applying a CBAM block \cite{woo2018cbam} on the features of the original image, and the final convolution is activated using a Sigmoid function.

\paragraph{\bf eye3 Team.}
This team implements the network architecture by Restormer \cite{zamir2022restormer}.
Instead of exploring new architecture for under-display camera restoration, they attempt to mine the potential of the existing method to advance this field. Following the scheme of \cite{zamir2022restormer}, they train an Restormer in a progressive fashion with $L_1$ loss for $300k$ iterations, with $92k$, $64k$, $48k$, $36k$, $36k$, and $24k$ iterations for patch size 128, 160, 192, 256, 320, and 384, respectively. Then they fix the patch size to $384 \times 384$ and use a mask loss strategy \cite{wu2021train} to train the model for another $36k$ iterations, where saturated pixels are masked out. After that, inspired by self-training \cite{xie2020self}, they add Gaussian noise to the well-trained model and repeat the training process.

\paragraph{\bf FMS Lab Team.}
Dual Branch Wavelet Net (DBWN) \cite{liu2018multi} that is proposed to restore the images degraded in an under-display camera imaging system. The overall network structure is similar to \cite{panikkasseril2020transform}, where the network branches restore the high and low spatial frequency components of the image separately.

\paragraph{\bf EDLC2004 Team.}
This team adopted TransWeather \cite{valanarasu2022transweather} model to deal with the image restoration problem for UDC images.
Specifically, they combine different loss functions including $L_1$ loss and $L2$ loss.
They train the model from scratch and it took approximately 17 hours with two 2080Ti GPUs.

\begin{figure}[t]
    \centering
     \includegraphics[width=0.8\textwidth]{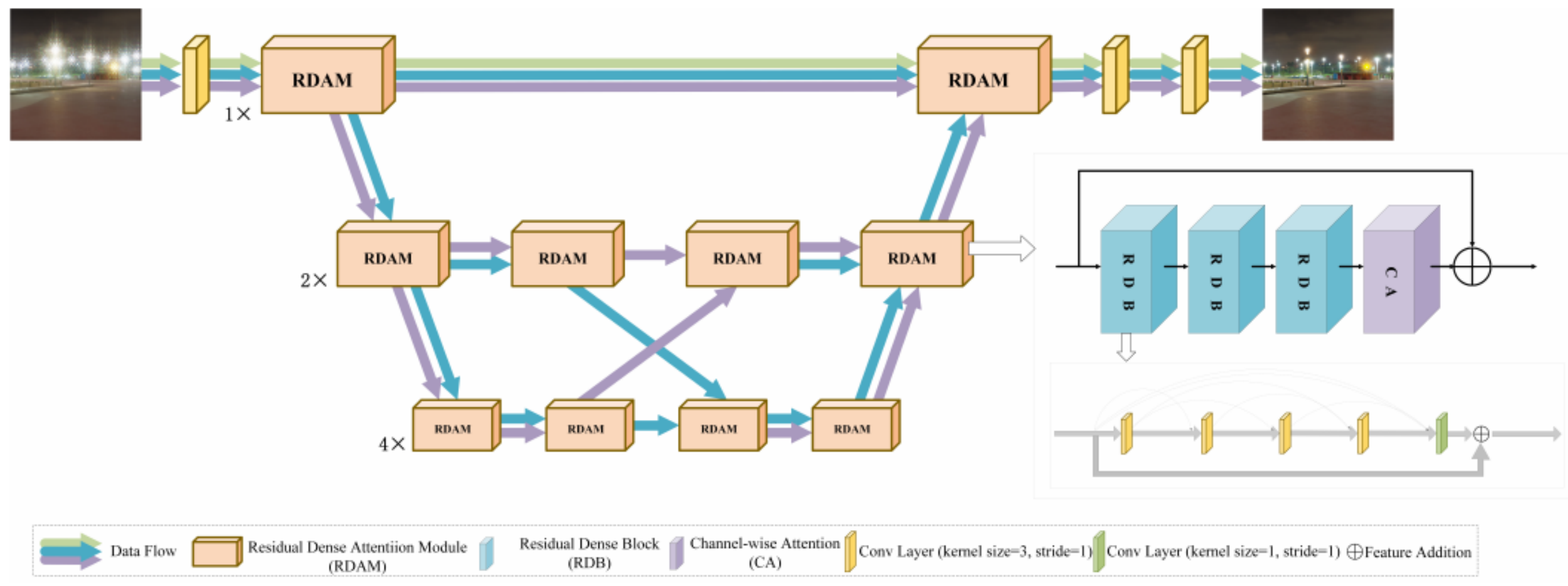}
    \caption{The network architecture proposed by SAU-LCFC Team.}
    \label{fig:sau_lcfc}
    \vspace{-0.7cm}
\end{figure}

\paragraph{\bf SAU\_LCFC Team.}
The team proposes an Hourglass-Structured Fusion Network (HSF-Net) as shown in Figure~\ref{fig:sau_lcfc}. They start from a coarse-scale stream, and then gradually repeat top-down and bottom-up processing to form more coarse-to-fine scale streams one by one. In each stream, the residual dense attention modules (RDAMs) are introduced as the basic component to deeply learn the features of under-display camera images. Each RDAM contains $3$ dense residual blocks (RDBs) in series with channel-wise attention (CA). In each RDB, the first 4 convolutional layers are adopted to elevate the number of feature maps, while the last convolutional layer is employed to aggregate feature maps. The growth rate of RDB is set to 16.

\section{Conclusions}
In this report, we review and summarize the methods and results of MIPI 2022 challenge on Under-Display Camera Image Restoration.
All the proposed methods are based on deep networks and most of them share a similar U-shape backbone to boost performance.

\noindent\textbf{Acknowledgements.}
We thank Shanghai Artificial Intelligence Laboratory, Sony, and Nanyang Technological University to sponsor this MIPI 2022 challenge. We thank all the organizers and all the participants for their great work.

\bibliographystyle{splncs04}
\bibliography{egbib}

\begin{thebibliography}{10}
\providecommand{\url}[1]{\texttt{#1}}
\providecommand{\urlprefix}{URL }
\providecommand{\doi}[1]{https://doi.org/#1}

\bibitem{abuolaim2021ntire}
Abuolaim, A., Timofte, R., Brown, M.S.: Ntire 2021 challenge for defocus
  deblurring using dual-pixel images: Methods and results. In: Proceedings of
  the IEEE/CVF Conference on Computer Vision and Pattern Recognition. pp.
  578--587 (2021)

\bibitem{chen2022simple}
Chen, L., Chu, X., Zhang, X., Sun, J.: Simple baselines for image restoration.
  arXiv preprint arXiv:2204.04676  (2022)

\bibitem{chen2017sca}
Chen, L., Zhang, H., Xiao, J., Nie, L., Shao, J., Liu, W., Chua, T.S.: Sca-cnn:
  Spatial and channel-wise attention in convolutional networks for image
  captioning. In: Proceedings of the IEEE/CVF Conference on Computer Vision and
  Pattern Recognition. pp. 5659--5667 (2017)

\bibitem{chen2021hdrunet}
Chen, X., Liu, Y., Zhang, Z., Qiao, Y., Dong, C.: Hdrunet: Single image hdr
  reconstruction with denoising and dequantization. In: Proceedings of the
  IEEE/CVF Conference on Computer Vision and Pattern Recognition. pp. 354--363
  (2021)

\bibitem{chi2020fast}
Chi, L., Jiang, B., Mu, Y.: Fast fourier convolution. Advances in Neural
  Information Processing Systems  \textbf{33},  4479--4488 (2020)

\bibitem{cho2021rethinking}
Cho, S.J., Ji, S.W., Hong, J.P., Jung, S.W., Ko, S.J.: Rethinking
  coarse-to-fine approach in single image deblurring. In: Proceedings of the
  IEEE/CVF International Conference on Computer Vision. pp. 4641--4650 (2021)

\bibitem{conde2022model}
Conde, M.V., McDonagh, S., Maggioni, M., Leonardis, A., P{\'e}rez-Pellitero,
  E.: Model-based image signal processors via learnable dictionaries. In:
  Proceedings of the AAAI Conference on Artificial Intelligence. vol.~36, pp.
  481--489 (2022)

\bibitem{dong2020deep}
Dong, J., Roth, S., Schiele, B.: Deep wiener deconvolution: Wiener meets deep
  learning for image deblurring. Advances in Neural Information Processing
  Systems  \textbf{33},  1048--1059 (2020)

\bibitem{feng2021removing}
Feng, R., Li, C., Chen, H., Li, S., Loy, C.C., Gu, J.: Removing diffraction
  image artifacts in under-display camera via dynamic skip connection networks.
  In: Proceedings of the IEEE/CVF Conference on Computer Vision and Pattern
  Recognition (2021)

\bibitem{huang2017densely}
Huang, G., Liu, Z., Van Der~Maaten, L., Weinberger, K.Q.: Densely connected
  convolutional networks. In: Proceedings of the IEEE/CVF Conference on
  Computer Vision and Pattern Recognition. pp. 4700--4708 (2017)

\bibitem{huang2021deep}
Huang, T., Dong, W., Yuan, X., Wu, J., Shi, G.: Deep gaussian scale mixture
  prior for spectral compressive imaging. In: Proceedings of the IEEE/CVF
  Conference on Computer Vision and Pattern Recognition. pp. 16216--16225
  (2021)

\bibitem{hui2019lightweight}
Hui, Z., Gao, X., Yang, Y., Wang, X.: Lightweight image super-resolution with
  information multi-distillation network. In: Proceedings of the 27th acm
  international conference on multimedia. pp. 2024--2032 (2019)

\bibitem{johnson2016perceptual}
Johnson, J., Alahi, A., Fei-Fei, L.: Perceptual losses for real-time style
  transfer and super-resolution. In: European Conference on Computer Vision.
  pp. 694--711. Springer (2016)

\bibitem{kingma2014adam}
Kingma, D.P., Ba, J.: Adam: A method for stochastic optimization. In: 3rd
  International Conference on Learning Representations, {ICLR} 2015, San Diego,
  CA, USA, May 7-9, 2015, Conference Track Proceedings (2015)

\bibitem{koh2022bnudc}
Koh, J., Lee, J., Yoon, S.: Bnudc: A two-branched deep neural network for
  restoring images from under-display cameras. In: Proceedings of the IEEE/CVF
  Conference on Computer Vision and Pattern Recognition. pp. 1950--1959 (2022)

\bibitem{liang2021swinir}
Liang, J., Cao, J., Sun, G., Zhang, K., Van~Gool, L., Timofte, R.: Swinir:
  Image restoration using swin transformer. In: Proceedings of the IEEE/CVF
  International Conference on Computer Vision. pp. 1833--1844 (2021)

\bibitem{liu2018multi}
Liu, P., Zhang, H., Zhang, K., Lin, L., Zuo, W.: Multi-level wavelet-cnn for
  image restoration. In: Proceedings of the IEEE Conference on Computer Vision
  and Pattern Recognition Workshops. pp. 773--782 (2018)

\bibitem{liu2018intriguing}
Liu, R., Lehman, J., Molino, P., Petroski~Such, F., Frank, E., Sergeev, A.,
  Yosinski, J.: An intriguing failing of convolutional neural networks and the
  coordconv solution. Advances in neural information processing systems
  \textbf{31} (2018)

\bibitem{liu2020single}
Liu, Y.L., Lai, W.S., Chen, Y.S., Kao, Y.L., Yang, M.H., Chuang, Y.Y., Huang,
  J.B.: Single-image hdr reconstruction by learning to reverse the camera
  pipeline. In: Proceedings of the IEEE/CVF Conference on Computer Vision and
  Pattern Recognition. pp. 1651--1660 (2020)

\bibitem{mao2016image}
Mao, X., Shen, C., Yang, Y.B.: Image restoration using very deep convolutional
  encoder-decoder networks with symmetric skip connections. Advances in neural
  information processing systems  \textbf{29} (2016)

\bibitem{nah2021ntire}
Nah, S., Son, S., Lee, S., Timofte, R., Lee, K.M.: Ntire 2021 challenge on
  image deblurring. In: Proceedings of the IEEE/CVF Conference on Computer
  Vision and Pattern Recognition. pp. 149--165 (2021)

\bibitem{panikkasseril2020transform}
Panikkasseril~Sethumadhavan, H., Puthussery, D., Kuriakose, M.,
  Charangatt~Victor, J.: Transform domain pyramidal dilated convolution
  networks for restoration of under display camera images. In: European
  Conference on Computer Vision. pp. 364--378. Springer (2020)

\bibitem{reinhard2002photographic}
Reinhard, E., Stark, M., Shirley, P., Ferwerda, J.: Photographic tone
  reproduction for digital images. In: Proceedings of the 29th annual
  conference on Computer graphics and interactive techniques. pp. 267--276
  (2002)

\bibitem{ronneberger2015u}
Ronneberger, O., Fischer, P., Brox, T.: U-net: Convolutional networks for
  biomedical image segmentation. In: International Conference on Medical Image
  Computing and Computer-Assisted Intervention. pp. 234--241. Springer (2015)

\bibitem{sun2019deep}
Sun, K., Xiao, B., Liu, D., Wang, J.: Deep high-resolution representation
  learning for human pose estimation. In: Proceedings of the IEEE/CVF
  conference on Computer Vision and Pattern Recognition. pp. 5693--5703 (2019)

\bibitem{timofte2016seven}
Timofte, R., Rothe, R., Van~Gool, L.: Seven ways to improve example-based
  single image super resolution. In: Proceedings of the IEEE/CVF conference on
  Computer Vision and Pattern Recognition. pp. 1865--1873 (2016)

\bibitem{valanarasu2022transweather}
Valanarasu, J.M.J., Yasarla, R., Patel, V.M.: Transweather: Transformer-based
  restoration of images degraded by adverse weather conditions. In: Proceedings
  of the IEEE/CVF Conference on Computer Vision and Pattern Recognition. pp.
  2353--2363 (2022)

\bibitem{wang2018esrgan}
Wang, X., Yu, K., Wu, S., Gu, J., Liu, Y., Dong, C., Qiao, Y., Change~Loy, C.:
  Esrgan: Enhanced super-resolution generative adversarial networks. In:
  Proceedings of the European Conference on Computer Vision Workshops. pp.
  63--79 (2018)

\bibitem{woo2018cbam}
Woo, S., Park, J., Lee, J.Y., Kweon, I.S.: Cbam: Convolutional block attention
  module. In: Proceedings of the European Conference on Computer Vision. pp.
  3--19 (2018)

\bibitem{wu2021train}
Wu, Y., He, Q., Xue, T., Garg, R., Chen, J., Veeraraghavan, A., Barron, J.T.:
  How to train neural networks for flare removal. In: Proceedings of the
  IEEE/CVF International Conference on Computer Vision. pp. 2239--2247 (2021)

\bibitem{xia2020basis}
Xia, Z., Perazzi, F., Gharbi, M., Sunkavalli, K., Chakrabarti, A.: Basis
  prediction networks for effective burst denoising with large kernels. In:
  Proceedings of the IEEE/CVF Conference on Computer Vision and Pattern
  Recognition. pp. 11844--11853 (2020)

\bibitem{xie2020self}
Xie, Q., Luong, M.T., Hovy, E., Le, Q.V.: Self-training with noisy student
  improves imagenet classification. In: Proceedings of the IEEE/CVF Conference
  on Computer Vision and Pattern Recognition. pp. 10687--10698 (2020)

\bibitem{zamir2022restormer}
Zamir, S.W., Arora, A., Khan, S., Hayat, M., Khan, F.S., Yang, M.H.: Restormer:
  Efficient transformer for high-resolution image restoration. In: Proceedings
  of the IEEE/CVF Conference on Computer Vision and Pattern Recognition. pp.
  5728--5739 (2022)

\bibitem{zamir2021multi}
Zamir, S.W., Arora, A., Khan, S., Hayat, M., Khan, F.S., Yang, M.H., Shao, L.:
  Multi-stage progressive image restoration. In: Proceedings of the IEEE/CVF
  Conference on Computer Vision and Pattern Recognition. pp. 14821--14831
  (2021)

\bibitem{zamir2022learning}
Zamir, S.W., Arora, A., Khan, S., Hayat, M., Khan, F.S., Yang, M.H., Shao, L.:
  Learning enriched features for fast image restoration and enhancement. arXiv
  preprint arXiv:2205.01649  (2022)

\bibitem{Zamir2022MIRNetv2}
Zamir, S.W., Arora, A., Khan, S., Hayat, M., Khan, F.S., Yang, M.H., Shao, L.:
  Learning enriched features for fast image restoration and enhancement. IEEE
  Transactions on Pattern Analysis and Machine Intelligence  (2022)

\bibitem{zhang2018unreasonable}
Zhang, R., Isola, P., Efros, A.A., Shechtman, E., Wang, O.: The unreasonable
  effectiveness of deep features as a perceptual metric. In: Proceedings of the
  IEEE/CVF Conference on Computer Vision and Pattern Recognition. pp. 586--595
  (2018)

\bibitem{zhang2018image}
Zhang, Y., Li, K., Li, K., Wang, L., Zhong, B., Fu, Y.: Image super-resolution
  using very deep residual channel attention networks. In: Proceedings of the
  European Conference on Computer Vision. pp. 286--301 (2018)

\bibitem{zhang2018residual}
Zhang, Y., Tian, Y., Kong, Y., Zhong, B., Fu, Y.: Residual dense network for
  image super-resolution. In: Proceedings of the IEEE/CVF Conference on
  Computer Vision and Pattern Recognition. pp. 2472--2481 (2018)

\bibitem{zhang2022frc}
Zhang, Z., Zheng, H., Hong, R., Fan, J., Yang, Y., Yan, S.: {FRC-Net: A Simple
  Yet Effective Architecture for Low-Light Image Enhancement}  (5 2022).
  \doi{10.36227/techrxiv.19771120.v2}

\bibitem{zhou2020udc}
Zhou, Y., Kwan, M., Tolentino, K., Emerton, N., Lim, S., Large, T., Fu, L.,
  Pan, Z., Li, B., Yang, Q., et~al.: Udc 2020 challenge on image restoration of
  under-display camera: Methods and results. In: European Conference on
  Computer Vision. pp. 337--351. Springer (2020)

\bibitem{zhou2021image}
Zhou, Y., Ren, D., Emerton, N., Lim, S., Large, T.: Image restoration for
  under-display camera. In: Proceedings of the IEEE/CVF Conference on Computer
  Vision and Pattern Recognition. pp. 9179--9188 (2021)

\end{thebibliography}
\clearpage

\appendix
\section{Teams and Affiliations}
\label{append:teams}

\subsection*{USTC\_WXYZ}
{\bf Title:} Enhanced Coarse-to-Fine Network for Restoring Images From Under-Display Cameras\\
{\bf Members:}\\
Yurui Zhu$^1$ (\href{mailto:zyr@mail.ustc.edu.cn}{zyr@mail.ustc.edu.cn})\\
Xi Wang$^1$\quad Xueyang Fu$^1$\quad Xiaowei Hu$^2$\\
{\bf Affiliations:}\\
$^1$ University of Science and Technology of China\\
$^2$ Shanghai AI Laboratory

\subsection*{XPixel Group}
{\bf Title:} UDC-UNet: Under-Display Camera Image Reconstruction via U-shape Dynamic Network\\
{\bf Members:}\\
Jinfan Hu$^1$ (\href{mailto:jf.hu1@siat.ac.cn}{jf.hu1@siat.ac.cn})\\
Xina Liu$^1$\quad Xiangyu Chen$^{1,2,3}$\quad Chao Dong$^{1,2}$\\
{\bf Affiliations:}\\
$^1$ Shenzhen Institutes of Advanced Technology, Chinese Academy of Sciences\\
$^2$ Shanghai AI Laboratory\\
$^3$ University of Macau

\subsection*{SRC-B}
{\bf Title:} MRNet: Multi-Refinement Network for Images Restoration on Under-display Camera\\
{\bf Members:}\\
Dafeng Zhang (\href{mailto:dfeng.zhang@samsung.com}{dfeng.zhang@samsung.com})\\
Feiyu Huang\quad Shizhuo Liu\quad Xiaobing Wang\quad Zhezhu Jin\\
{\bf Affiliations:}\\
Samsung Research China, Beijing

\subsection*{MIALGO}
{\bf Title:} Residual Dense Network Based on Multi-Resolution Feature Fusion\\
{\bf Members:}\\
Xuhao Jiang (\href{mailto:jiangxuhao@xiaomi.com}{jiangxuhao@xiaomi.com})\\
Guangqi Shao\quad Xiaotao Wang\quad Lei lei\\
{\bf Affiliations:}\\
Xiaomi, Beijing

\subsection*{LVGroup\_HFUT}
{\bf Title:} Towards lightweight and real-time under-display camera image restoration\\
{\bf Members:}\\
Zhao Zhang (\href{mailto:cszzhang@gmail.com}{cszzhang@gmail.com})\\
Suiyi Zhao\quad Huan Zheng\quad Yangcheng Gao\quad Yanyan Wei\quad Jiahuan Ren\\
{\bf Affiliations:}\\
Hefei University of Technology

\subsection*{GSM}
{\bf Title:} Deep Gaussian Scale Mixture Prior for UDC Image Restoration\\
{\bf Members:}\\
Tao Huang (\href{mailto:thuang_666@stu.xidian.edu.cn}{thuang\_666@stu.xidian.edu.cn})\\
Zhenxuan Fang\quad Mengluan Huang\quad Junwei Xu\\
{\bf Affiliations:}\\
School of Artificial Intelligence, Xidian University

\subsection*{Y2C}
{\bf Title:} N/A\\
{\bf Members:}\\
Yong Zhang$^1$ (\href{mailto:yongzhang@whu.edu.cn}{yongzhang@whu.edu.cn})\\
Yuechi Yang$^1$\quad Qidi Shu$^2$\quad Zhiwen Yang$^1$\quad Shaocong Li$^1$\\
{\bf Affiliations:}\\
$^1$ School of Remote Sensing and Information Engineering, Wuhan University\\
$^2$ State Key Laboratory of Information Engineering in Surveying, Mapping and Remote Sensing, Wuhan University

\subsection*{VIDAR}
{\bf Title:} Deep Gaussian Scale Mixture Prior for UDC Image Restoration\\
{\bf Members:}\\
Mingde Yao (\href{mailto:mdyao@mail.ustc.edu.cn}{mdyao@mail.ustc.edu.cn})\\
Ruikang Xu\quad Yuanshen Guan\quad Jie Huang\quad Zhiwei Xiong\\
{\bf Affiliations:}\\
University of Science and Technology of China

\subsection*{IILLab}
{\bf Title:} Wiener Guided Coupled Dynamic Filter Network\\
{\bf Members:}\\
Hangyan Zhu (\href{mailto:zhuhy0309@gmail.com}{zhuhy0309@gmail.com})\\
Ming Liu\quad Shaohui Liu\quad Wangmeng Zuo\\
{\bf Affiliations:}\\
Harbin Institute of Technology

\subsection*{jzsherlock}
{\bf Title:} U-RRDBNet for UDC Image Restoration\\
{\bf Members:}\\
Zhuang Jia (\href{mailto:jiazhuang@xiaomi.com}{jiazhuang@xiaomi.com})\\
{\bf Affiliations:}\\
Xiaomi

\subsection*{Namecantbenull}
{\bf Title:} NAFNet\\
{\bf Members:}\\
Binbin SONG (\href{mailto:yb97426@um.edu.mo}{yb97426@um.edu.mo})\\
{\bf Affiliations:}\\
University of Macau

\subsection*{MeVision}
{\bf Title:} Two-Branched Network for Image Restoration of Under-display Camera\\
{\bf Members:}\\
Ziqi Song (\href{mailto:songziqi@mail.ustc.edu.cn}{songziqi@mail.ustc.edu.cn})\\
Guiting Mao\quad Ben Hou\quad Zhimou Liu\quad Yi Ke\quad Dengpei Ouyang\quad Dekui Han\\
{\bf Affiliations:}\\
Changsha Research Institute of mining and metallurgy

\subsection*{BIVLab}
{\bf Title:} Self-Guided Coarse-to-Fine Network for Progressive Under-Display Camera Image Restoration\\
{\bf Members:}\\
Jinghao Zhang (\href{mailto:jhaozhang@mail.ustc.edu.cn}{jhaozhang@mail.ustc.edu.cn})\\
QiZhu\quad Naishan Zheng\quad Feng Zhao\\
{\bf Affiliations:}\\
University of Science and Technology of China

\subsection*{RushRushRush}
{\bf Title:} N/A\\
{\bf Members:}\\
Wu Jin (\href{mailto:monster_w@tju.edu.cn}{monster\_w@tju.edu.cn})\\
{\bf Affiliations:}\\
Tianjin University

\subsection*{JMU-CVLab}
{\bf Title:} Lightweight Dual-branch UDC Blind Image Restoration\\
{\bf Members:}\\
Marcos Conde$^1$ (\href{mailto:marcos.conde-osorio@uni-wuerzburg.de}{marcos.conde-osorio@uni-wuerzburg.de})\\
Sabari Nathan$^2$\quad Radu Timofte$^1$\\
{\bf Affiliations:}\\
$^1$ University of Wurzburg, Computer Vision Lab, Germany\\
$^2$ Couger Inc., Japan

\subsection*{eye3}
{\bf Title:} N/A\\
{\bf Members:}\\
Tianyi Xu (\href{mailto:tillyandbamboo@163.com}{tillyandbamboo@163.com})\\
Jun Xu\\
{\bf Affiliations:}\\
School of Statistics and Data Science, Nankai University

\subsection*{FMS Lab}
{\bf Title:} Dual Branch Wavelet Net (DBWN)\\
{\bf Members:}\\
Hrishikesh P.S.$^1$ (\href{mailto:hrishikesh@foundingminds.com}{hrishikesh@foundingminds.com})\\
Densen Puthussery$^1$\quad Jiji C.V.$^2$\\
{\bf Affiliations:}\\
$^1$ Founding Minds Software\\
$^2$ Department of Electronics and Communication SRM University

\subsection*{EDLC2004}
{\bf Title:} N/A\\
{\bf Members:}\\
Jiang Biao (\href{mailto:jiangb22@m.fudan.edu.cn}{jiangb22@m.fudan.edu.cn})\\
Ding Yuhan\quad Li WanZhang\quad Feng Xiaoyue\quad Chen Sijing\quad Zhong Tianheng\\
{\bf Affiliations:}\\
Fudan University

\subsection*{SAU\_LCFC}
{\bf Title:} Hourglass-Structured Fusion Network (HSF-Net)\\
{\bf Members:}\\
Jiyang Lu$^1$ (\href{mailto:lujiyang1@stu.sau.edu.cn}{lujiyang1@stu.sau.edu.cn})\\
Hongming Chen$^1$\quad Zhentao Fan$^1$\quad Xiang Chen$^2$\\
{\bf Affiliations:}\\
$^1$ Shenyang Aerospace University\\
$^2$ Nanjing University of Science and Technology

\end{sloppypar}
\end{document}